\documentclass[pdflatex,sn-mathphys-num]{sn-jnl}
\usepackage{graphicx}%
\usepackage{multirow}%
\usepackage{amsmath,amssymb,amsfonts}%
\usepackage{amsthm}%
\usepackage{mathrsfs}%
\usepackage[title]{appendix}%
\usepackage{xcolor}%
\usepackage{textcomp}%
\usepackage{manyfoot}%
\usepackage{booktabs}
\usepackage{algorithm}%
\usepackage{algorithmicx}%
\usepackage{algpseudocode}%
\usepackage{listings}%
\usepackage{tabularx}%
\usepackage{tikz}%
\theoremstyle{thmstyleone}%
\theoremstyle{thmstyletwo}%
\theoremstyle{thmstylethree}%
\newcolumntype{Y}{>{\centering\arraybackslash}X}
\definecolor{Orange}{cmyk}{0, 0.5, 1, 0}     
\definecolor{SkyBlue}{cmyk}{0.8, 0, 0, 0}
\definecolor{BluishGreen}{cmyk}{0.97, 0, 0.75, 0}
\definecolor{Yellow}{cmyk}{0.1, 0.05, 0.9, 0}
\definecolor{Blue}{cmyk}{1, 0.5, 0, 0}
\definecolor{Vermillion}{cmyk}{0, 0.8, 1, 0}
\definecolor{ReddishPurple}{cmyk}{0.1, 0.7, 0, 0}
\definecolor{BlueP}{rgb}{0     , 0.4470, 0.7410}
\definecolor{RedP}{rgb}{0.8500, 0.3250, 0.0980}
\definecolor{YellowP}{rgb}{0.9290, 0.6940, 0.1250}
\definecolor{PurpleP}{rgb}{0.4940, 0.1840, 0.5560}
\definecolor{GreenP}{rgb}{0.4660, 0.6740, 0.1880}
\definecolor{SkyBlueP}{rgb}{0.3010, 0.7450, 0.9330}
\definecolor{DarkRedP}{rgb}{0.6350, 0.0780, 0.1840}
\DeclareRobustCommand\blackline{\tikz[baseline=-0.6ex]\draw[thick] (0,0)--(0.54,0);}
\DeclareRobustCommand\blackdashed{\tikz[baseline=-0.6ex]\draw[thick,dashed] (0,0)--(0.54,0);}
\DeclareRobustCommand\blackdashdotted{\tikz[baseline=-0.6ex]\draw[thick,dashdotted] (0,0)--(0.54,0);}
\DeclareRobustCommand\blueline{\tikz[baseline=-0.6ex]\draw[BlueP,thick] (0,0)--(0.54,0);}
\DeclareRobustCommand\redline{\tikz[baseline=-0.6ex]\draw[RedP,thick] (0,0)--(0.54,0);}
\DeclareRobustCommand\yellowline{\tikz[baseline=-0.6ex]\draw[YellowP,thick] (0,0)--(0.54,0);}
\DeclareRobustCommand\purpleline{\tikz[baseline=-0.6ex]\draw[PurpleP,thick] (0,0)--(0.54,0);}
\DeclareRobustCommand\greenline{\tikz[baseline=-0.6ex]\draw[GreenP,thick] (0,0)--(0.54,0);}
\DeclareRobustCommand\bluedashed{\tikz[baseline=-0.6ex]\draw[BlueP,thick,dashed] (0,0)--(0.54,0);}
\DeclareRobustCommand\bluedashdotted{\tikz[baseline=-0.6ex]\draw[BlueP,thick,dashdotted] (0,0)--(0.54,0);}
\DeclareRobustCommand\bluedotted{\tikz[baseline=-0.6ex]\draw[BlueP,thick,dotted] (0,0)--(0.54,0);}
\DeclareRobustCommand\reddashed{\tikz[baseline=-0.6ex]\draw[RedP,thick,dashed] (0,0)--(0.54,0);}
\DeclareRobustCommand\yellowdashed{\tikz[baseline=-0.6ex]\draw[YellowP,thick,dashed] (0,0)--(0.54,0);}
\DeclareRobustCommand\greendotted{\tikz[baseline=-0.6ex]\draw[GreenP,thick,dotted] (0,0)--(0.54,0);}

\begin{document}
%
\title[]{Relaminarization of turbulent pipe flow induced by streamwise traveling wave wall transpiration and its scaling}
%
\author[1]{\fnm{Christian} \sur{Bauer}}\email{christian.bauer@dlr.de}
\author[1,2]{\fnm{Claus} \sur{Wagner}}
\affil[1]{\orgdiv{Institute of Aerodynamics and Flow Technology}, \orgname{German Aerospace Center (DLR)}, \orgaddress{\street{Bunsenstrasse 10}, \city{G\"ottingen}, \postcode{37073}, \country{Germany}}}
\affil[2]{\orgdiv{Institute of Thermodynamics and Fluid Mechanics}, \orgname{Technische Universit\"at Ilmenau}, \orgaddress{\street{Am Helmholtzring 1}, \city{Ilmenau}, \postcode{98684}, \country{Germany}}}
%
\abstract{
In technical applications, fluids are often pumped through pipes, creating turbulent flows with high Reynolds numbers. Since more than 90\% of the energy required to pump the fluids is dissipated by turbulence near the wall, relaminarization of such flows is a way to save a significant amount of energy.
In this respect, streamwise traveling waves of wall blowing and suction have been used to relaminarize turbulent pipe flow at a low friction Reynolds number of $\mathrm{Re}_\tau=110$, considerably reducing friction losses and energy consumption. Whether this also applies to higher Reynolds numbers is the subject of this work. We demonstrate that streamwise traveling waves of wall blowing and suction applied to initially turbulent pipe flow can trigger relaminarization up to friction Reynolds numbers of $\mathrm{Re}_\tau=720$.
Furthermore, we perform a parametric study comprising both upstream traveling waves (UTWs, $c<0$) and downstream traveling waves (DTWs, $c>0$) by varying the traveling wave amplitude $a$, celerity $c$, and wavelength $\lambda$ at $\mathrm{Re}_\tau=180$ and $\mathrm{Re}_\tau=360$. This study is conducted  in order to investigate the scaling of the maximum drag reduction and of the net energy saving rate in direct numerical simulations.
Consistent with channel flow studies in the literature, we found that UTWs destabilize the flow, while generating sublaminar drag due to the pumping effect.
However, only low-speed UTWs with large amplitudes were discovered to decisively reduce energy consumption.
For DTWs, a large range of wave parameters lead to a conspicuous drag reduction. Nevertheless, only a subgroup of these wave parameters are associated with siginificant net energy savings,
i.e. $0.067U_{c,lam} \lesssim a \lesssim0.1U_{c,lam}$, $c \approx U_{c,lam}$, and $\lambda\approx 360\delta_\nu$, independent of the Reynolds number.
Here, $U_{c,lam}=1/2\mathrm{Re}_\tau u_\tau$ is the centerline velocity of the corresponding laminar flow, and $\delta_\nu = \nu/u_\tau$ is the viscous length scale.
In the cases of relaminarization, it takes more than $65D/u_\tau$ for the flow to accelerate to its terminal velocity. Meanwhile, the turbulent kinetic energy damps to nearly zero within $3D/u_\tau$, independent of the Reynolds number. 
In particular, the decay of the turbulent kinetic energy and its components, as well as the Reynolds shear stress, can be well approximated by an exponential decay function.
The relaminarized flow pattern scales in mixed coordinates due to traveling wave-induced half-vortical structures that scale in viscous units and are superimposed with a laminar velocity profile which scales with the pipe radius.
These structures effectively reduce the pipe cross-section, resulting in a smaller mean velocity profile for the relaminarization cases than for the corresponding laminar Hagen-Poiseuille profile.
Nonetheless, more than $97\%$ of the theoretically achievable drag reduction rate is obtained within the Reynolds number range of $180\le\mathrm{Re}_\tau\le540$.
In brief, the observed scaling relations enable the prediction of traveling wave-induced relaminarization of turbulent pipe flow at higher Reynolds numbers than before.\par

}
%
%
%
\maketitle
%
\section{Introduction}
\label{sec:intro}
Pipe flows are vital to numerous engineering applications, such as oil and gas pipelines. However, a large fraction of the energy needed to push fluid through pipes is dissipated near the pipe wall due to turbulence.
One way to save energy is to laminarize the turbulent flow, which recovers a substantial amount of the dissipated energy.
Specifically, the friction drag coefficient of laminar pipe flow is $C_{f,lam}=16/\mathrm{Re}_b$, whereas the friction drag coefficient of turbulent pipe flow can be estimated as 
\begin{equation}
C_{f,turb}\approx0.0707\mathrm{Re}_b^{-0.24}.\label{eq:cffit}
\end{equation}
This illustrates that the relaminarization of a turbulent pipe flow at bulk Reynolds numbers of $\mathrm{Re}_b=u_bD/\nu \approx 25000$ can reduce friction losses by more than 90 percent.
Here, $u_b$ is the bulk velocity, $D$ is the pipe diameter, and $\nu$ is the kinematic viscosity.
Note that equation~(\ref{eq:cffit}) corresponds to the relation introduced by \citet{Blasius1913}, fitted by recent direct numerical simulation (DNS) data~\cite{ElKhoury2013,Ahn2015,Bauer2017,Bauer2021,Pirozzoli2021,Yao2023} and predicting the friction drag coefficient with an accuracy of more than 98\% up to bulk Reynolds numbers of $\mathrm{Re}_b = 240000$.
Recently, \citet{Kuhnen2018,Kuhnen2018a,Kuhnen2019} 
demonstrated that modifying the streamwise velocity profile of an initially fully-developed turbulent pipe flow using different control methods can result in relaminarization at high Reynolds numbers ($\mathrm{Re}_b\approx 10^5$).\par
Another method of reducing friction drag in wall-bounded turbulence involves directly manipulating velocity fluctuations and consequently the Reynolds shear stress.
According to \citet{Fukagata2002}, the skin friction coefficient in fully-developed turbulent pipe flow can be decomposed into a laminar and a turbulent contribution, i.e.,
\begin{equation}
C_f = \frac{16}{\mathrm{Re}_b} + 16\int_{0}^{1}2r\langle u^\prime_r u^\prime_z \rangle r dr,\label{eq:fik}
\end{equation}
where velocities are normalized by twice the bulk velocity $u_b$ and the radial coordinate is normalized by the pipe radius $R$.
The second term in equation~(\ref{eq:fik})---also known as the FIK identity---corresponds to the weighted integration of the Reynolds shear stress $\langle u^\prime_ru^\prime_z\rangle$.
Therefore, to minimize the drag in wall-bounded turbulence, control mechanisms often aim to minimize the Reynolds shear stress.
Using DNSs of turbulent plane channel flow at a bulk Reynolds number of $Re_b=2000$, \citet{Min2006} showed that upstream traveling waves (UTW) generated by a zero-net mass-flux surface blowing and suction as wall-normal velocity boundary conditions can decrease the skin friction drag below that of the corresponding laminar profile by inducing a negative Reynolds shear stress in equation~(\ref{eq:fik}).
Moreover, \citet{Hoepffner2009} demonstrated that traveling waves of wall blowing and suction applied to a channel flow without a mean pressure gradient induce a bulk flow in the opposite direction of the traveling waves.
This mechanism is also known as the pumping effect and is responsible for the drag reduction in channel flows with UTWs.
However, although UTWs have the potential to realize sublaminar drag, \citet{Bewley2009} and \citet{Fukagata2009} proved mathematically that the total energy required to drive and control the main flow can not be lower than the energy required for driving the corresponding laminar flow.
\citet{Moarref2010} predicted that the onset of turbulence in channel flow could be controlled by a downstream traveling wave (DTW) with the proper speed and frequency, whereas UTWs destabilize the flow.  
Using DNSs of channel flow at a low Reynolds number ($\mathrm{Re}_\tau=u_\tau h/\nu=63$), \citet{Lieu2010} confirmed these predictions, showing that DTWs can relaminarize fully-developed turbulent channel flow and that UTWs promote turbulence, even when the base flow is laminar.
Above, $u_\tau$ is the friction velocity and $h$ is the channel's half-height.
\par
%
In addition to wall blowing and suction, streamwise traveling waves can be induced by wall deformation.
\citet{Nakanishi2012} demonstrated that DTWs of wall deformation applied to turbulent channel flow with a constant flow rate (CFR) corresponding to a bulk Reynolds number of $\mathrm{Re}_b=5300$ cause relaminarization of the flow, accompanied by a  69\% drag reduction and 65\% net energy savings.
Several studies~\cite{Frohnapfel2012,Hasegawa2014,Gatti2018} have pointed out that, for flow control cases under a constant flow rate (CFR) condition, drag reduction is defined by the decrease in the resultant pressure gradient and thus in the friction Reynolds number $\mathrm{Re}_\tau$. In contrast, for cases with a constant pressure gradient (CPG), drag reduction corresponds to an increase in flow rate and therefore in the bulk Reynolds number $\mathrm{Re}_b$.
In the instance of turbulent channel flow under CPG condition, \citet{Nabae2020} performed a parametric study with DTW wall deformation at friction Reynolds numbers of $\mathrm{Re}_\tau=90,180,360$.
Unlike in the case of the CFR condition~\cite{Nakanishi2012}, \citet{Nabae2020} did not observe relaminarization of the flow, but rather drag reduction caused by negative periodic Reynolds shear stress, similar to UTWs of wall blowing and suction.
Based on their parametric study, \citet{Nabae2020} proposed a semi-empirical formula for the mean velocity profile that predicts 20\% drag reduction at practically high Reynolds numbers of $\mathrm{Re}_\tau=10^6$.
Recently, \citet{Nabae2025} investigated the drag reduction effect caused by streamwise traveling wave wall deformation in high-Reynolds number turbulent channel flow using large-eddy simulation and  confirmed the aforementioned semi-empirical formula up to friction Reynolds numbers of $\mathrm{Re}_\tau=3240$.
As for a spatially developing, compressible, turbulent boundary layer flow, \citet{Albers2024} achieved a drag reduction of over 50\% using streamwise traveling wave wall deformation.\par
%
To assess the parameter range in which drag reduction or relaminarization occurs in blowing/suction-controlled turbulent channel flow under the CPG condition, \citet{Mamori2014} carried out a large number of DNSs at $\mathrm{Re}_\tau=110$ and $\mathrm{Re}_\tau=300$ varying the amplitude $a$, wavelength $\lambda$, and celerity $c$ of the traveling wave boundary condition. 
Thus, they determined the effective parameters to obtain relaminarization under DTW conditions at both Reynolds numbers, i.e.
$a>0.1 U_{b,lam}$, $c>1.5U_{b,lam}$, and $200 < \lambda^+ <500$. Here, $U_{b,lam}$ is the bulk velocity of the corresponding laminar flow.
An upper bound for the celerity and amplitude was not clearly assessed.
These wave parameters result in a displacement thickness ranging from 3 to 10 wall units, which leads to relaminarization.
Moreover, \citet{Mamori2014} showed that a drag reduction rate of 0.15 can be achieved with UTWs for $\mathrm{Re}_\tau=110$, depending on the wavelength.
Following this approach, \citet{Koganezawa2019} conducted a parametric study of DTW boundary conditions in turbulent pipe flow at $\mathrm{Re}_\tau=110$ using DNSs.
They exposed that, for $\mathrm{Re}_\tau=110$, DTWs can initially relaminarize turbulent pipe flow for $a>0.03U_{c,lam}$, $c>U_{c,lam}$, and $150 < \lambda^+<500$ where $U_{c,lam}=1/2\mathrm{Re}_\tau u_\tau$ is the centerline velocity of the corresponding laminar flow.
Note that, for laminar pipe flow, $U_{b,lam}=1/2U_{c,lam}$.
The above parametric studies~\cite{Mamori2014,Koganezawa2019} determine the range of parameters at which relaminarization occurs in DTW controlled internal wall-bounded flows, i.e. turbulent channel and pipe flow, under the CPG condition. 
They put forth that the wavelength $\lambda$ scales with the viscous length scale $\delta_\nu=\nu/u_\tau$. 
Additionally, they suggest that the celerity $c$ and the amplitude $a$ scale with the centerline velocity of the corresponding laminar flow $U_{c,lam}$. Notwithstanding, larger Reynolds number data are required to confirm these scalings.
Besides, to the author's knowledge, no turbulent pipe flow DNSs with UTW boundary conditions have been performed so far.
The present study investigates the effects of drag reduction and relaminarization induced by UTW and DTW boundary conditions on the wall-normal velocity component in turbulent pipe flow under the CPG condition at Reynolds numbers ranging from $180 \le \mathrm{Re}_\tau \le 720$ and for various values of the parameters $a$, $\lambda$, and $c$.
In addition to scaling of the drag reduction rate, this study examines the scaling of the net energy savings rate, as the latter is an important quantity for determining the energy efficiency of a flow control method.
For selected flow cases, the temporal evolution and scaling of statistical quantities, such as the mean velocity and the turbulent kinetic energy profiles,incl are analyzed.
%
%
\section{Flow geometry and numerical methodology}
\label{sec:methodology}
\begin{figure} 
\centering
\includegraphics[]{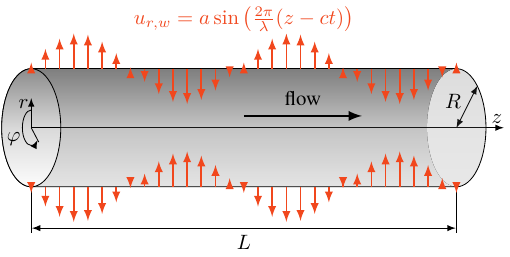}
	\caption{Pipe geometry involving a cylindrical coordinate system, where $z$ is the axial, $\varphi$ the azimuthal and $r$ the radial coordinate. $L$ is the pipe length and $R$ the pipe radius. At the pipe wall the wall-normal velocity is set to $u_{r,w}=a\sin(2\pi/\lambda(z-ct))$}
\label{fig:geom}
\end{figure}

The governing equations in case of a pressure-driven incompressible flow of a Newtonian fluid in a smooth pipe are the incompressible Navier-Stokes equations in the following dimensionless form
\begin{eqnarray}
\frac{\partial \vec u}{\partial t} + \vec u \cdot \nabla \vec u + \nabla p &=& \frac{1}{\mathrm{Re}_\tau}\nabla^2 \vec u, \label{eq:mom}  \\ 
\nabla \cdot \vec u &=& 0,
\label{eq:mass}
\end{eqnarray}
where $\mathrm{Re}_\tau=u_\tau R / \nu$ is the friction Reynolds number, based on the pipe radius $R$. Equation~(\ref{eq:mom}) is integrated in time using a fourth-order finite volume method based on a leapfrog-Euler time integration scheme~\cite{Feldmann2012,Bauer2017}. 
The flow geometry, which is a smooth pipe with length $L$ and radius $R$, as shown in Fig.~\ref{fig:geom}, is discretized via staggered grids in a cylindrical coordinate system.
While the no-slip boundary condition is applied to the axial and azimuthal velocity component at the wall, i.e. $u_{z,w}=u_{\varphi,w}=0$, the wall-normal velocity component at the wall is set to a traveling-wave-like velocity profile,
\begin{equation}
	u_{r,w}=u_r(z,\varphi,r=R)=a\sin\left(\frac{2\pi}{\lambda}(z-ct)\right), \label{eq:waveeq}
\end{equation}
with $a$ the amplitude, $\lambda$ the wavelength, and $c$ the celerity of the traveling wave.
Based on the friction Reynolds number and the parameters $a$, $\lambda$, and $c$, a set of DNSs has been performed, see Table~\ref{tab:cases}.
In addition, uncontrolled reference cases with an impermeable wall boundary condition, i.e. $a=0$, have been carried out for each Reynolds number. 
Note that statistical quantities obtained from the reference cases are denoted with the subscript $0$.
\begin{table} 
\caption{Summary of the turbulent pipe flow simulation cases: $\mathrm{Re}_\tau= u_\tau R/\nu$ is the friction Reynolds number and \# the number of simulation cases per friction Reynolds number. $N_z , N_{\phi}$, and $N_r$ are the number of grid points with respect to the axial, azimuthal, and radial direction, respectively. $\Delta{z^+}$, streamwise grid spacing; $R^+\Delta{\varphi}$ azimuthal grid spacing at the wall; $\Delta{r^+}$, minimal and maximal radial grid spacing. $a$, $\lambda$, and $c$ are the amplitude, the wavelength, and the celerity of the traveling wave boundary condition. The superscript $+$ denotes spacings normalized in wall units using the viscous length $\delta_\nu = \nu/u_\tau$, with friction velocity $u_\tau$.}
\begin{tabularx}{\linewidth}{lYYYY} 
\hline\hline
$\mathrm{Re}_\tau$  &   180             &   360         &   540         &   720     \\
\#                  &   81             &   44          &   1           &     1      \\
$L/R$               &   20              &   14           &   14           &   14       \\
$N_z$               &   768             &   1024        &   1536        &   1536    \\
$N_\varphi$         &   256             &   512         &   1024        &   1024    \\
$N_r$               &   84              &   160         &   222         &   222     \\
$a^+$               &   $\hspace{1ex}4.5\dots27$    &   $\hspace{3ex}9\dots27$  &   24& 24          \\
$\lambda^+$         &   $\hspace{3ex}36\dots720$    & $\hspace{2ex}180\dots720$ &   360       & 504          \\
$c^+$               &   $-270\dots360$  & $-360\dots360$& 270 &  180         \\
$\Delta{z^+}$       &   $4.7$           &   $4.9$       &   $4.9$       &   $6.6$   \\
$R^+\Delta{\phi}$   &   $4.4$           &   $4.4$       &   $4.9$       &   $6.6$   \\
$\Delta r^+$        &   $0.31\dots4.4$  & $0.39\dots4.4$& $0.37\dots4.9$&$0.49\dots6.6$\\
\hline\hline
\label{tab:cases}
\end{tabularx}
\end{table}
In the following, the angle brackets indicate an average in space and/or time according to the subscript.
First, for monitoring the temporal evolution of the simulations, volume-averaged time series of a quantity $\phi$ are computed as
\begin{equation}
\langle \phi \rangle_{z \varphi r}(t)=\frac{1}{L\pi R^2}\int_{0}^{L}\int_{0}^{R}\int_{0}^{2\pi}\phi(z,\varphi,r,t)r d\varphi dr dz. \label{eq:avgzphir}
\end{equation}
Furthermore, due to the inhomogeneities of the flow with respect to the streamwise and radial direction as well as the transient behavior of the flow, the following azimuthal averaging is taken into account:
\begin{equation}
\langle \phi \rangle_{\varphi}(z,r,t)=\frac{1}{2\pi }\int_{0}^{2\pi}\phi(z,\varphi,r,t) d\varphi, \label{eq:avgphi}
\end{equation}
and instantaneous turbulent fluctuations are defined as the difference between an instantaneous quantity and its azimuthal mean, i.e.
\begin{equation}
\phi^\prime=\phi-\langle \phi \rangle_\varphi. \label{eq:fluc}
\end{equation}
In the event that the applied control leads to a new statistically stationary solution, an additional temporal averaging is applied to a flow quantity $\phi$ to achieve statistically-converged quantities, i.e.
\begin{equation}
\langle \phi \rangle_{t} =\frac{1}{T}\int_{t_0}^{t_0+T}  \phi  dt, \label{eq:avgzphirt}
\end{equation}
with $t_0$ being the start of the time interval $T$, where the flow behaves statistically stationary.
Consequently, a volume- and time-averaged quantity is denoted as $\langle \phi \rangle_{z\varphi r t}=\langle\langle\phi\rangle_{z \varphi r}\rangle_t$.
%
Moreover, to distinguish between the $z$-averaged component, the periodic component, and the turbulent component of a flow variable $\phi$, the three-component decomposition by~\citet{Hussain1970} is introduced,
\begin{equation}
\phi(z,\varphi,r,t) = \langle \phi \rangle_{z\varphi t}(r) +\tilde{\phi}(\tilde{z},r) + \phi^{\prime\prime}(z,\varphi,r,t),
\label{eq:threedecomp}
\end{equation}
with the spatio-temporal average,
\begin{equation}
\langle \phi \rangle_{z \varphi t}(r)=\frac{1}{2\pi L T}\int_{t_0}^{t_0+T}\int_{0}^{L}\int_{0}^{2\pi}\phi(z,\varphi,r,t) d\varphi dz dt, \label{eq:avgzphit}
\end{equation}
the periodic component,
\begin{equation}
\tilde{\phi}(\tilde{z},r) = \langle\phi \rangle_{N_{\phi_z}\varphi t}(\tilde{z},r)-\langle \phi \rangle_{z\varphi t}(r), 
\label{eq:phitilde}
\end{equation}
and the turbulent component $\phi^{\prime\prime}(z,\varphi,r,t)$. 
In equation~(\ref{eq:phitilde}), the pase-average of $\phi$ is computed as
\begin{equation}
\langle\phi \rangle_{N_{\phi_z}\varphi t}(\tilde{z},r) = \frac{1}{2\pi TN_{\phi_z}}\sum_{z\in N_{\phi_z}}\left(\int_{t_0}^{t_0+T}\int_{0}^{2\pi}\phi(z,\varphi,r,t) d\varphi dt\right),
\label{eq:phaseavg}
\end{equation}
where $N_{\phi_z}=L/\lambda$ is the number of streamwise locations of the same phase and $\tilde{z}=\lambda n+(z-ct)$ is the streamwise coordinate in phase space, with the integer $n$ and $0 \le \tilde{z}\le \lambda$.\par
For the present set of simulations, the pressure gradient is given as a constant input parameter, and the volume flow rate, i.e. the bulk velocity, 
\begin{equation}
u_b = \langle u_z \rangle_{z \varphi r},  \label{eq:ubulk}
\end{equation}
is an output parameter. 
Moreover, the skin friction drag is computed from the bulk velocity as follows:
\begin{equation}
C_f = \frac{2\tau_w}{\rho u_b^2}=\frac{2}{u_b^{+2}},  \label{eq:skinfriction}
\end{equation}
where $\tau_w$ is the wall shear stress, $\rho$ is the fluid density, and the superscript $+$ denotes normalization in wall units, i.e. normalization with the friction velocity $u_\tau$ for velocities and the viscous length scale $\delta_\nu$ for length scales.
Then, the drag reduction rate $R_D$ is computed as follows:
\begin{equation}
R_D = \frac{C_{f0}-C_f}{C_{f0}},  \label{eq:rd}
\end{equation}
with the skin friction coefficient of the uncontrolled case $C_{f0}$ at either the same friction Reynolds number or the same bulk Reynolds number as the uncontrolled flow.
In the former case $C_{f0}$ is obtained by additional temporal averaging over equation~(\ref{eq:skinfriction}) for the uncontrolled flow.
To assess the efficiency of the control method, we introduce the net energy savings rate,
\begin{equation}
S = \frac{W_{p0}-(W_p+W_a)}{W_{p0}},  \label{eq:s}
\end{equation}
with $W_p$($W_{p0}$) the driving power of the (un)controlled flow and $W_a$ the actuation power required by the control, viz.
\begin{eqnarray}
W_p &=& -\left\langle \frac{\partial p}{\partial z} \right\rangle L u_b \frac{\pi D^2}{4} = \tau_w u_b \pi D L, \label{eq:wp}\\
W_a &=& \int_0^{L}\int_0^{2\pi} \left(\frac{1}{2}u^3_{r,w}+p_wu_{r,w}\right)Rd\varphi dz\notag\\
&=& \langle p_w u_{r,w}\rangle_{z\varphi} \pi D L. \label{eq:wa}
\end{eqnarray}
Inserting expression (\ref{eq:wp}) and (\ref{eq:wa}) into (\ref{eq:s})  and reformulating in wall units result in
\begin{equation}
S = \frac{(u_{b,0}^+)^{-2}-\left((u_{b}^+)^{-2}+\langle p^+_w u^+_{r,w}\rangle_{z\varphi}(u_b^+)^{-3}\right)}{(u_{b,0}^+)^{-2}}, \label{eq:s2}
\end{equation}
where $u^+_{b,0}$ corresponds to the bulk velocity of the uncontrolled flow in wall units at the same bulk Reynolds number as the controlled flow. The latter is estimated using the relation
\begin{equation}
u_{b,0}^+ \approx 5.319 \mathrm{Re}_b^{0.12}, \label{eq:ubfit}
\end{equation}
see~\citet{Bauer2021}.
It is worth mentioning, that additional energy losses depending on the mechanical efficiency of the control device might further reduce the net energy savings in real flow applications.\par
%
%
%
%
\section{Results}
\label{sec:results}
%
\subsection{Drag reduction and net energy saving}
\label{ss:drag}
\begin{figure} 
\centering
\includegraphics[]{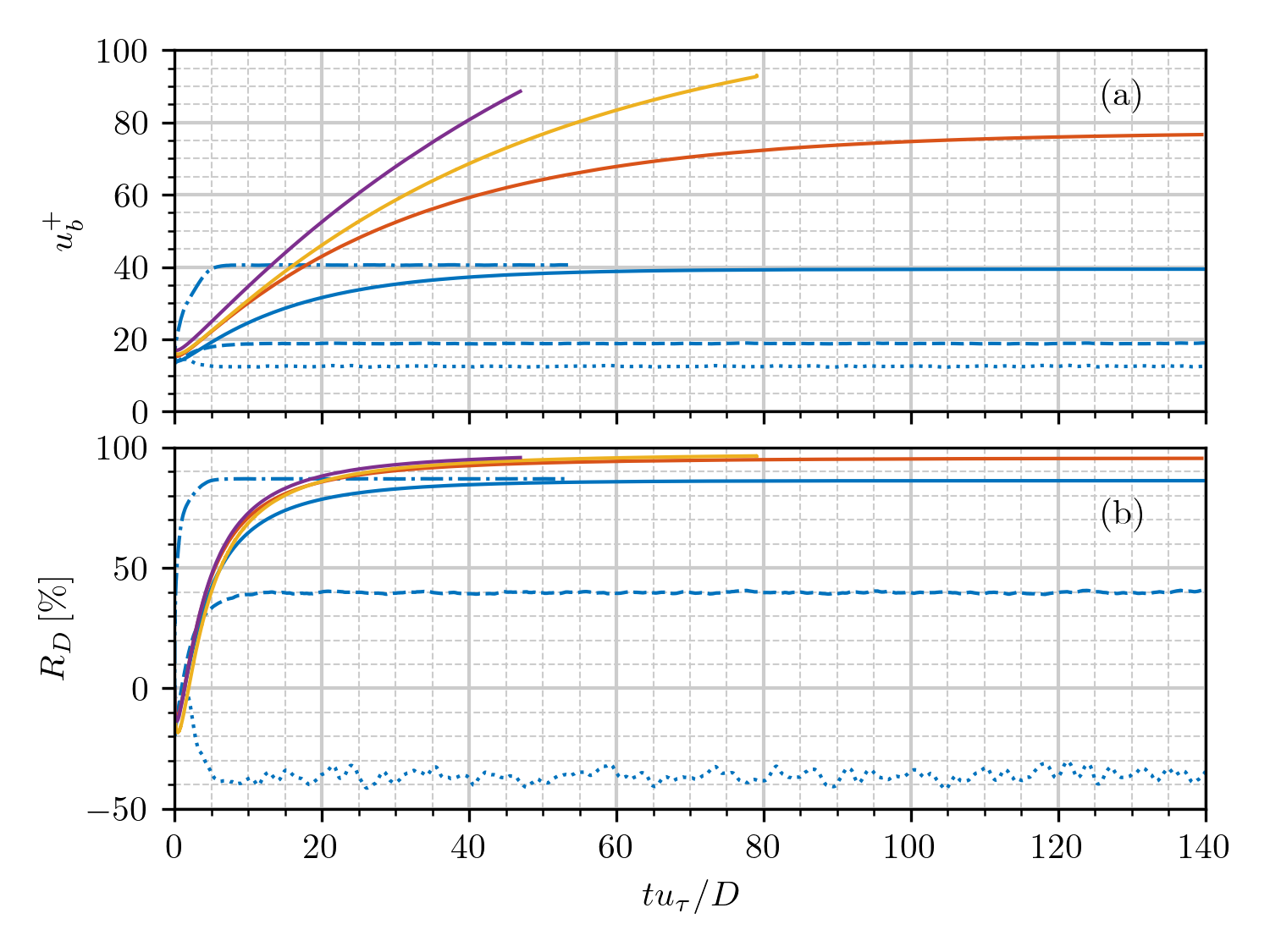}
\caption{
Time series of the bulk velocity $u^+_b$ (a) and drag reduction rate $R_D$ (b) for selected flow cases  at $\mathrm{Re}_\tau=180$ (blue lines), $\mathrm{Re}_\tau=360$ (red line), $\mathrm{Re}_\tau=540$ (yellow line), and $\mathrm{Re}_\tau=720$ (purple line). DTWs with(without) relaminarization, solid(dotted) lines; standing wave, dashed line; UTW, dashed-dotted line. In detail:\blueline, case 1; \bluedashed, case 2; \bluedashdotted, case 3; \bluedotted, case 4; \redline, case 5; \yellowline, case 6; \purpleline, case 7; see Table~\ref{tab:selcases}.
}
\label{fig:rd_vs_t}
\end{figure}

To evaluate the effectiveness of the control, the drag reduction rate $R_D$ is computed during the simulations by solving equation (\ref{eq:rd}) on-the-fly.
Fig.~\ref{fig:rd_vs_t} shows the time series of the bulk velocity and the drag reduction rate for the selected flow cases presented in Table~\ref{tab:selcases}. 
Here, the uncontrolled skin friction drag $C_{f0}$ is obtained from direct numerical simulations with the corresponding friction Reynolds numbers.

\begin{table} 
\caption{Selected flow cases: $\mathrm{Re}_\tau=$ is the friction Reynolds number, $a^+$, $\lambda^+$, and $c^+$ are the amplitude, the wavelength, and the celerity of the traveling wave boundary condition in wall units.}
\begin{tabularx}{\linewidth}{lYYYYY} 
\hline\hline
case    & $\mathrm{Re}_\tau$    & $a^+$     & $\lambda^+$   & $c^+$     &line\\
\hline
1       &  180                  &  6        & 360           & 90        &\blueline\\
2       &  180                  &  6        & 360           & 0         &\bluedashed\\
3       &  180                  &  27       & 360           & -9        &\bluedashdotted\\
4       &  180                  &  4.5      & 360           & 90        &\bluedotted\\
5       &  360                  &  13.5     & 360           & 180       &\redline\\
6       &  540                  &  24.0     & 360           & 270       &\yellowline\\
7       &  720                  &  24.0     & 504           & 180        &\purpleline\\
\hline\hline
\label{tab:selcases}
\end{tabularx}
\end{table}
%
%
As indicated by the blue lines in Fig.~\ref{fig:rd_vs_t} for a friction Reynolds number of $\mathrm{Re}_\tau=180$, the terminal bulk velocity $u_b$ and, consequently, the drag reduction rate $R_D$ depend strongly on the wave parameters $a^+$, $\lambda^+$, and $c^+$.
For a wavelength of $\lambda^+=360$ and a celerity of $c^+=90$, the bulk velocity increases to $u_b^+=39.4$ for an amplitude of $a^+=6$ (case 1, blue line in Fig.~\ref{fig:rd_vs_t}a). In contrast, the bulk velocity decreases to $u_b^+=12.6$ for an amplitude of $a^+=4.5$ (case 4, dotted blue line in Fig.~\ref{fig:rd_vs_t}a).
This highlights the sensitivity of the achieved friction drag regarding the changes in the control parameters.
Notably, using UTWs with a celerity of $c^+=-9$ (case~3, with $a^+=6$, and $\lambda^+=360$) leads to a sudden increase of the bulk velocity to $u_b^+=40.6$, resulting in a significant drag reduction of approximately 87\% (dashed-dotted line in Fig.~\ref{fig:rd_vs_t}b).
Additionally, the case of a standing wave with $c=0$ (case 2, with $a^+=6$, and $\lambda^+=360$) causes an intermediate increase in terminal bulk velocity to $u_b^+=18.7$ and a subsequent drag reduction of about 39\% (dashed line in Fig.~\ref{fig:rd_vs_t}b).
As Reynolds numbers increase, cases 5, 6, and 7 correspond to cases with a significant increase in $u_b^+$ and $R_D$ for DTWs.
As we will substantiate later, in these cases, the reduction in friction drag coincides with a relaminarization of the flow.
However, Fig.~\ref{fig:rd_vs_t}(b) shows that the time period of turbulent pipe flow relaminarization, and corresponding drag decrease, scales with the friction velocity $u_\tau$ and the pipe diameter $D$ for DTWs. 
Specifically, the solid lines in Fig.~\ref{fig:rd_vs_t}(b) portray that the drag reduction rate converges after approximately 65 dimensionless time units, at which point $R_D$ reaches 99\% of its final value.\par
\begin{figure} 
\includegraphics[trim=11 0 17 0,clip]{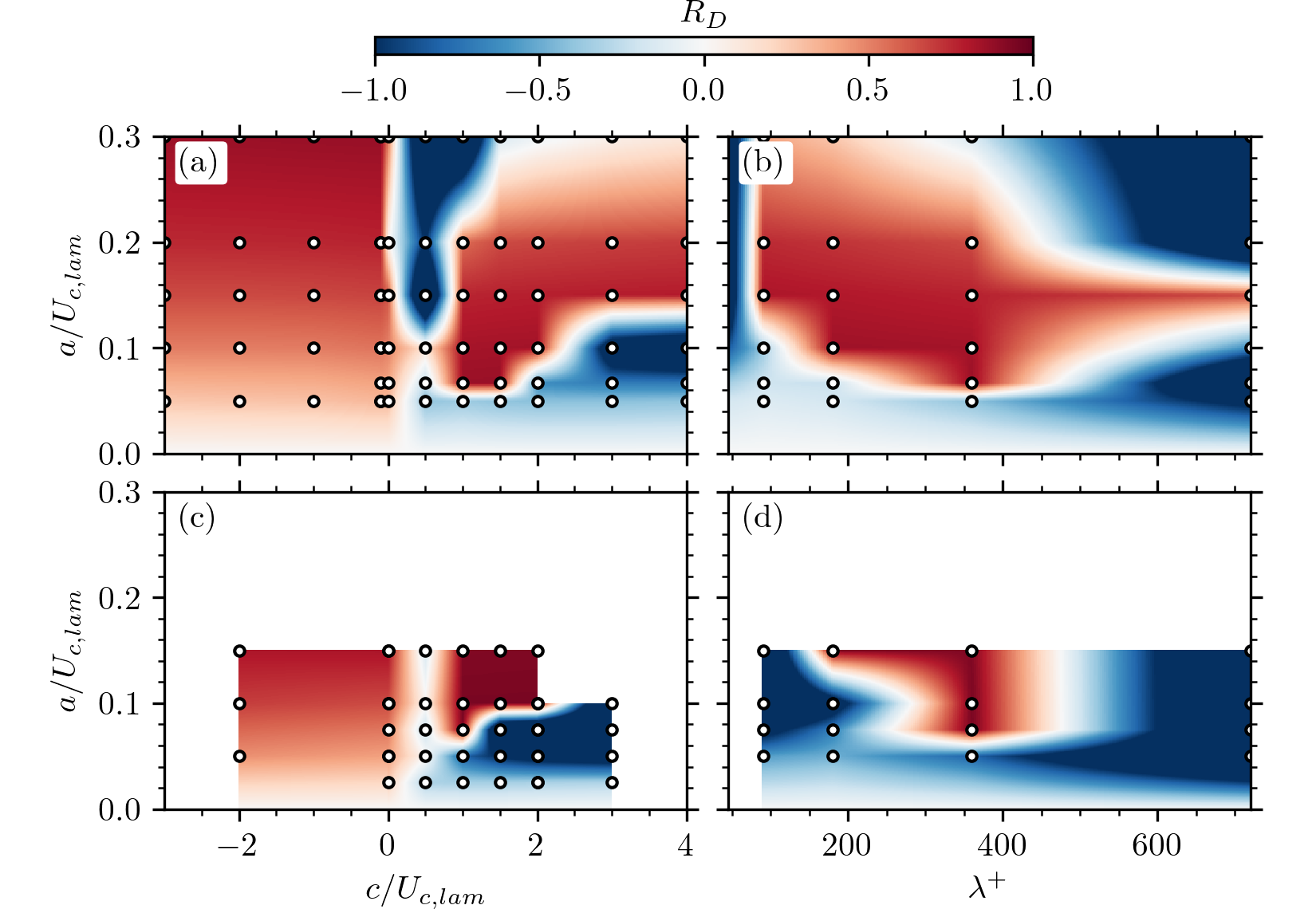}
\caption{
Drag reduction $R_D=(C_{f0}-C_f)/C_{f0}$ for $\mathrm{Re}_\tau=180$ (a,b) and $\mathrm{Re}_\tau=360$ (c,d). 
(a,c) $\lambda^+=360$; (b) $c/U_{c,lam}=1.5$, (d) $c/U_{c,lam}=1.0$.
White circles indicate simulation cases.
}
\label{fig:rd}
\end{figure}
\begin{figure} 
\centering
\includegraphics[trim=11 0 17 0,clip]{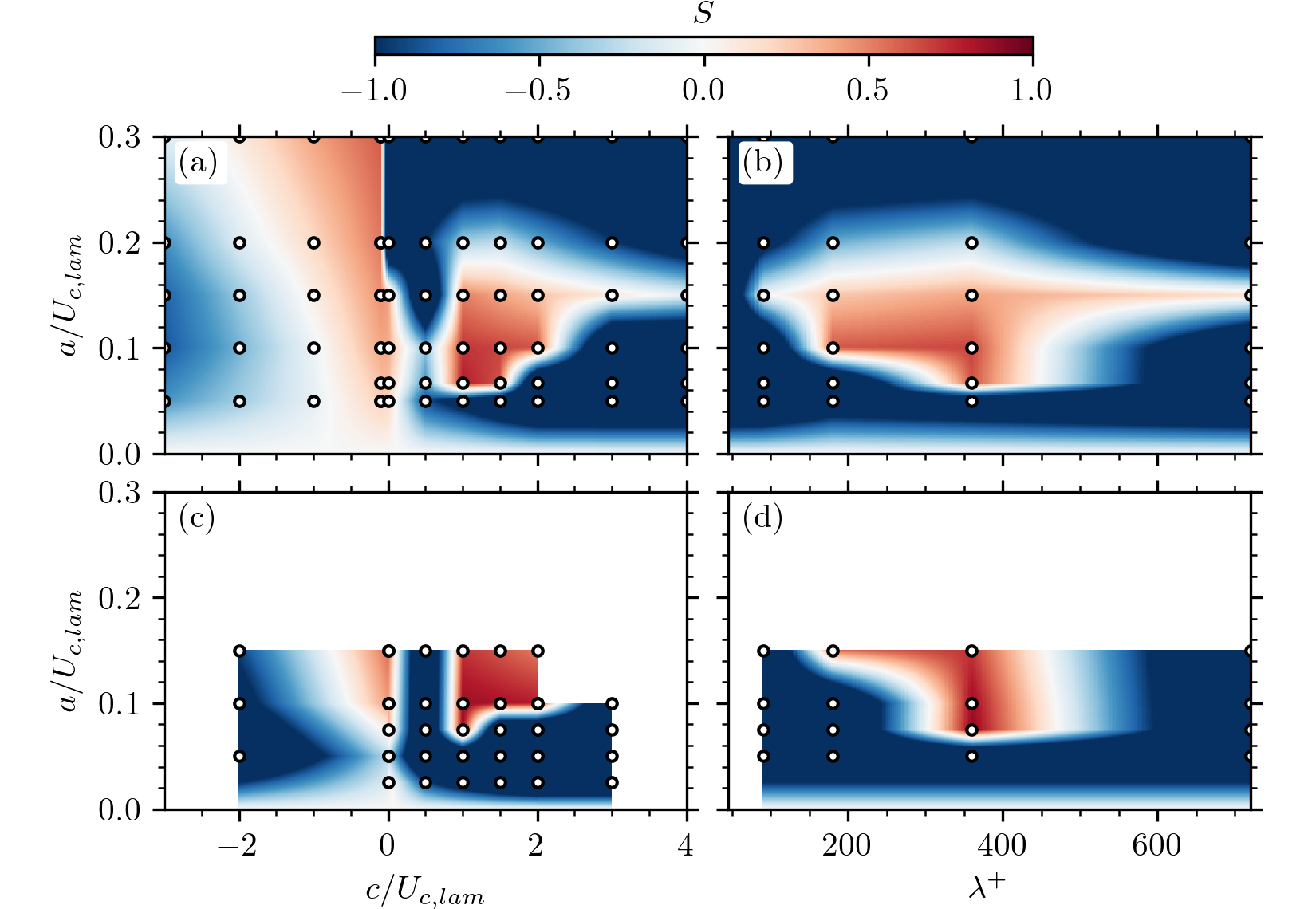}
\caption{
The net energy savings are given by $S=(W_{p0}-(W_p+W_a))/W_{p0}$ for $\mathrm{Re}_\tau=180$ (a,b) and $\mathrm{Re}_\tau=360$ (c,d). 
(a) and (c): $\lambda^+=360$; (b): $c/U_{c,lam}=1.5$; (d): $c/U_{c,lam}=1.0$.
White circles indicate simulation cases.
}
\label{fig:s}
\end{figure}
\begin{figure} 
\centering
\includegraphics[trim=5 0 5 0,clip,width=\linewidth]{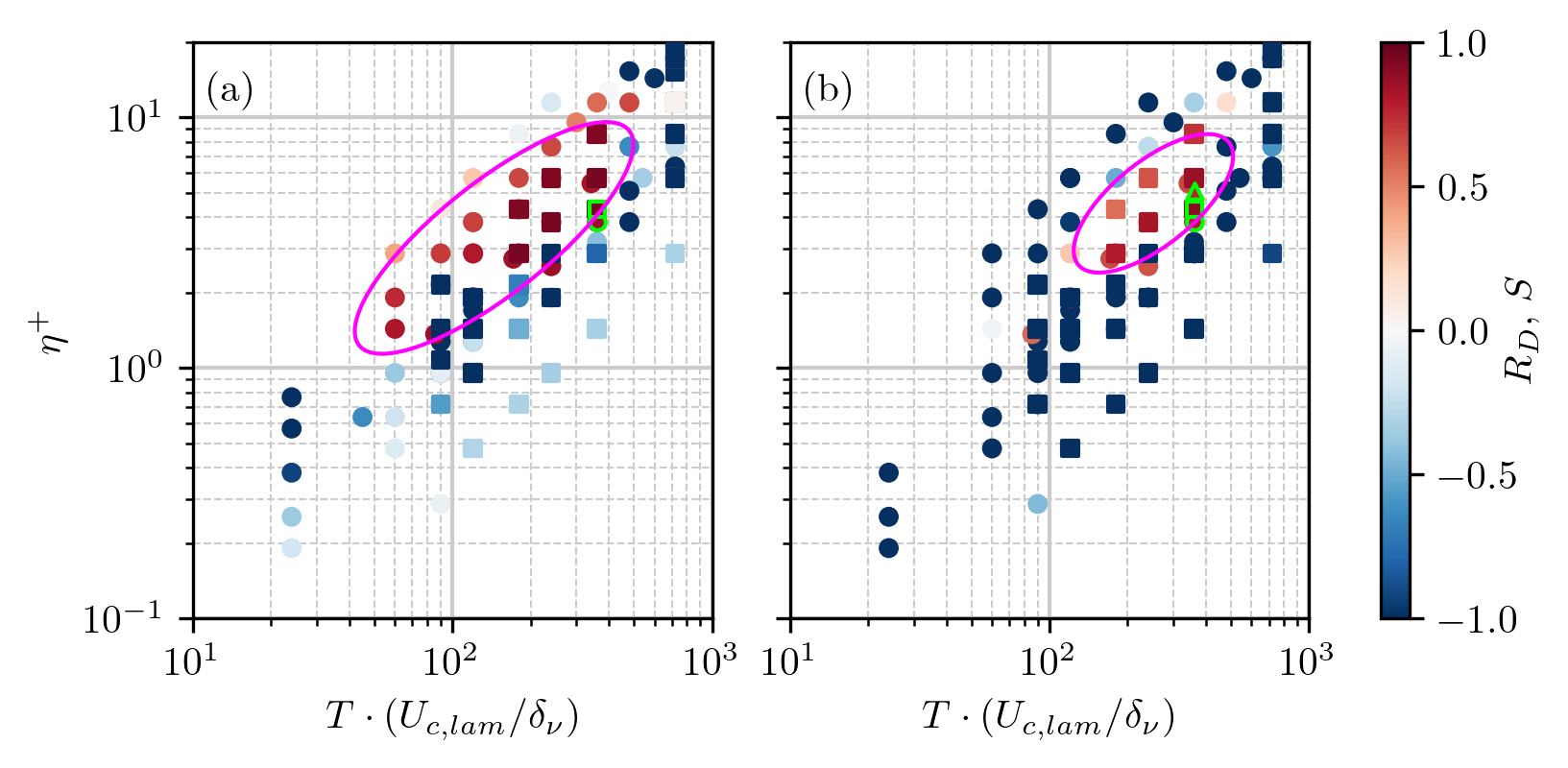}
\caption{
Drag reduction rate $R_D$ (a) and net energy saving rate $S$ (b) versus the effective layer thickness $\eta^+$ in wall units and the control period $T$ normalized by $\delta_\nu/U_{c,lam}$. Circles, $\mathrm{Re}_\tau=180$; squares, $\mathrm{Re}_\tau=360$; triangles, $\mathrm{Re}_\tau=540$.
The magenta line shows the effective parameter ranges for drag reduction (a) and net energy savings (b). 
Green outline corresponds to relaminarization cases~1,~5, and~6.
}
\label{fig:etat}
\end{figure}

To determine the control's effectiveness with respect to the wave parameters, Fig.~\ref{fig:rd} presents the drag reduction rate $R_D$ as pseudo-color images for friction Reynolds numbers of $\mathrm{Re}_\tau=180$ (a,b) and $\mathrm{Re}_\tau=360$ (c,d).
The drag reduction maps for a constant wavelength of $\lambda^+=360$ and a constant celerity of $c/U_{c,lam}=1.5$ at $\mathrm{Re}_\tau=180$ and $c/U_{c,lam}=1.0$ for $\mathrm{Re}_\tau=360$ are shown in Fig.~\ref{fig:rd}(a,c) and Fig.~\ref{fig:rd}(b,d), respectively.
Note that both the wave amplitude and celerity in Fig.~\ref{fig:rd} are normalized by the centerline velocity of the corresponding laminar flow, $U_{c,lam}=1/2\mathrm{Re}_\tau u_\tau $, in accordance with the literature~\cite{Mamori2014,Koganezawa2019}.
The pseudo-color maps in Fig.~\ref{fig:rd} are interpolated from the converged $R_D$ values of the 81 DNSs performed for $\mathrm{Re}_\tau=180$ and of the 44 simulations for $\mathrm{Re}_\tau=360$.
The DNSs are indicated by white circles in Fig.~\ref{fig:rd}.
In short, Fig.~\ref{fig:rd}(a,c) conveys that, for UTWs ($c<0$), the terminal drag reduction rate increases with the wave amplitude $a$ independently of celerity $c$ for both displayed Reynolds numbers.
For DTWs ($c>0$), however, a minimum wave amplitude ($a\gtrsim 0.07U_{c,lam}$) and celerity ($c>U_{c,lam}$) are necessary for drag reduction, which are nearly independent of the Reynolds number (Fig.~\ref{fig:rd}a,c).
Furthermore, several combinations of larger wave amplitudes and celerities lead to drag reduction up to $a=0.3U_{c,lam}$ and $c=4U_{c,lam}$. At these values, approximately 10\% of the friction drag is still reduced compared to the uncontrolled flow case.
Additionally, drag reduction is achieved at a constant celerity of $c=1.5U_{c,lam}$ at $\mathrm{Re}_\tau=180$ for a range of wavelengths $\lambda^+$ (Fig.~\ref{fig:rd}b).
For instance, substantial drag reduction is attained for DTWs with amplitude $a=0.15U_{c,lam}$ and wavelengths in the range $90\le\lambda^+\le720$.
By contrast, for the friction Reynolds number of $\mathrm{Re}_\tau=360$ presented in Fig.~\ref{fig:rd}(d), only the wavelength of $\lambda^+=360$ causes drag reduction for all considered cases except for $a=0.15U_{c,lam}$, where an additional wavelength of ($\lambda^+=180$) was found to also cause significant drag reduction.
Although the drag reduction rate is a valid measure of the method's effectiveness, it does not provide information about its efficiency.
Since efficiency is an even more important quantity for evaluating flow control methods, this paper discusses the net energy savings $S$, defined in equation (\ref{eq:s}).
Fig.~\ref{fig:s} depicts $S$ in the same planes that cut through the three-dimensional parameter space of $a$, $c$, and $\lambda$ as Fig.~\ref{fig:rd}.
At first glance, it is clear from Fig.~\ref{fig:s} and a comparison with Fig.~\ref{fig:rd} that the regions in the parameter space characterized by net energy savings are significantly smaller than the regions of drag reduction.
For UTW cases with $c \lesssim -2U_{c,lam}$, negative net energy savings are obtained, meaning more energy is consumed by the actuation of the flow control than is gained by the drag reduction.
For UTWs ($c<0$) and $\mathrm{Re}_\tau=180$, only cases with a low celerity ($c=-0.1U_{c,lam}$) result in significant net energy savings of up to 61\%, depending on the wave amplitude $a$ (Fig.~\ref{fig:s}a).
For standing waves ($c=0$), positive net energy savings are reached for wave amplitudes up to $a=0.15U_{c,lam}$ as shown in Fig.~\ref{fig:s}(a,b).
It should be noted that this value is the largest amplitude investigated for $\mathrm{Re}_\tau=360$, as can be seen in Fig.~\ref{fig:s}(b).
The maximum net energy savings rate for $c=0$ and $\lambda^+=360$ is approximately 50\% for $\mathrm{Re}_\tau=180$ and 55\% for $\mathrm{Re}_\tau=360$.
By comparison, the maximum net energy saving for DTWs in Fig.~\ref{fig:s} is located around $c/U_{c,lam}\approx1$, $0.067 \lesssim a/U_{c,lam}\lesssim 0.1$, and $\lambda^+\approx 360$ for both Reynolds numbers $\mathrm{Re}_\tau=180$ and $\mathrm{Re}_\tau=360$.
This is in good agreement with the wave parameters $c/U_{c,lam}\approx1$, $a/U_{c,lam}\approx 0.091$, and $\lambda^+\approx 345$ for the maximum drag reduction at $\mathrm{Re}_\tau=110$ reported by \citet{Koganezawa2019}.
To reduce the wave parameter space from three to two dimensions, we introduce the effective layer thickness,
\begin{equation}
\eta = \frac{a\lambda}{2\pi c},
\end{equation}
and the control period, 
\begin{equation}
T=\frac{\lambda}{c}.
\end{equation}
The effective layer thickness is obtained by integrating equation~(\ref{eq:waveeq}) over time~\cite{Mamori2014,Koganezawa2019}.
Fig.~\ref{fig:etat} shows the drag reduction rate $R_D$ (a) and the net energy saving rate $S$ (b) in the reduced parameter space of the control period $T$ and the effective layer thickness $\eta$. 
In accordance with the scaling of the wave parameters, the control period is normalized by $\delta_\nu/U_{c,lam}$ and the effective layer thickness is normalized by $\delta_\nu$.
As displayed in Fig.~\ref{fig:etat}(a), the effective layer thickness for drag reduction at $180 \le \mathrm{Re_\tau} \le 540$ is $1.4\lesssim \eta^+ \lesssim 11.5$, which is linearly coupled to the time period $60 \lesssim TU_{c,lam}/\delta_\nu \lesssim 480$. This corresponds well to the results reported by \citet{Koganezawa2019} for traveling-wave controlled pipe flow at $\mathrm{Re}_\tau=110$.
However, Fig.~\ref{fig:etat}(b) shows that only a small range of parameters ranges leads to net energy savings independent of the Reynolds number. This range is $3.8 \lesssim \eta^+ \lesssim 5.7$ and $T\approx360 \delta_\nu/U_{c,lam}$.
Table~\ref{tab:maxsaving} summarizes the maximum net energy savings and the corresponding drag reduction rate of the present study and that of \citet{Koganezawa2019}.
\begin{table} 
\caption{The maximum achievable net energy savings $S$ and drag reduction $R_D$ for each friction Reynolds number $\mathrm{Re}_\tau$ are listed together with $a/U_{c,lam}$ and $c/U_{c,lam}$, which are the traveling wave amplitude and celerity, respectively, normalized by the laminar centerline velocity $U_{c,lam}=1/2\mathrm{Re}_\tau u_\tau$. Furthermore, $\lambda^+$ is the traveling wave length in wall units, $\eta^+$ is the effective layer thickness in wall units, and $T$ is the control period. 
For $\mathrm{Re}_\tau=110$, the values are reported by \citet{Koganezawa2019}.
}
\begin{tabularx}{\linewidth}{lYYYYYYY} 
\hline\hline
$\mathrm{Re}_\tau$    & $a/U_{c,lam}$     & $\lambda^+$   & $c/U_{c,lam}$  &$\eta^+$ & $T{U_{c,lam}/\delta_\nu}$   &$S$ & $R_D$\\
\hline
110       &    0.091      &  345        & 1        &  5.0  & 345  &  0.51      & 0.60\\
180       &    0.067      &  360        & 1        &  3.8 &  360 &  0.79       & 0.82\\
360       &    0.075      &  360        & 1        &  4.3 &   360 & 0.90     & 0.93\\
540       &    0.089      &  360        & 1        &  5.1 &   360 & 0.88   & 0.95\\
\hline\hline
\label{tab:maxsaving}
\end{tabularx}
\end{table}
While the maximum attainable drag reduction rate increases with the Reynolds number, no clear trend is observable for the net energy savings rate.
In the present study, the maximum net energy savings rate of $S\approx90\%$ is achieved at $\mathrm{Re}_\tau=360$.\par
\subsection{Temporal evolution of the turbulent kinetic energy}
\label{ss:tke}
\begin{figure} 
\centering
\includegraphics[]{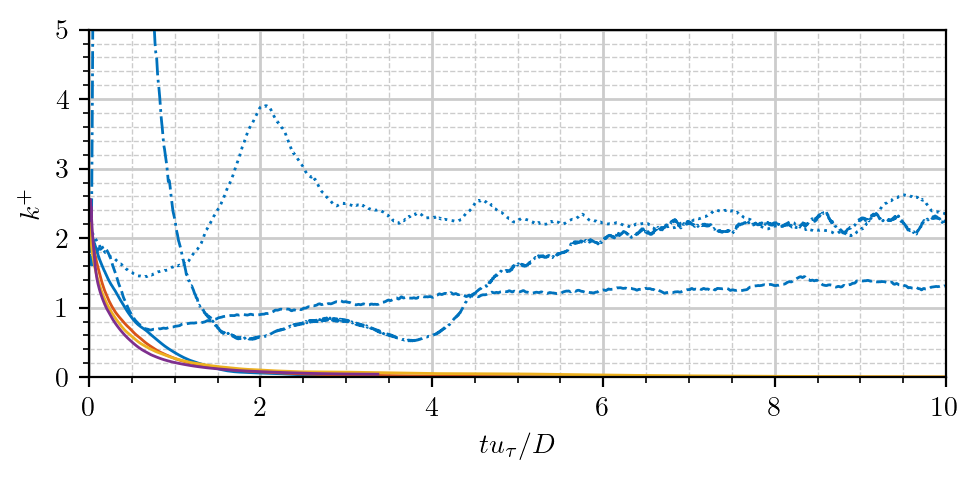}
\caption{
Time series of the turbulent kinetic energy $k^+$ for selected flow cases at $\mathrm{Re}_\tau=180$ (blue lines), $\mathrm{Re}_\tau=360$ (red line), $\mathrm{Re}_\tau=540$ (yellow line), and $\mathrm{Re}_\tau=720$ (purple line). DTWs with(without) relaminarization, solid(dotted) lines; standing wave, dashed line; UTW, dashed-dotted line. In detail: \blueline, case 1; \bluedashed, case 2; \bluedashdotted, case 3; \bluedotted, case 4; \redline, case 5; \yellowline, case 6; \purpleline, case 7; see Table~\ref{tab:selcases}.
}
\label{fig:tke_vs_t}
\end{figure}

\begin{figure} 
\centering
\includegraphics[]{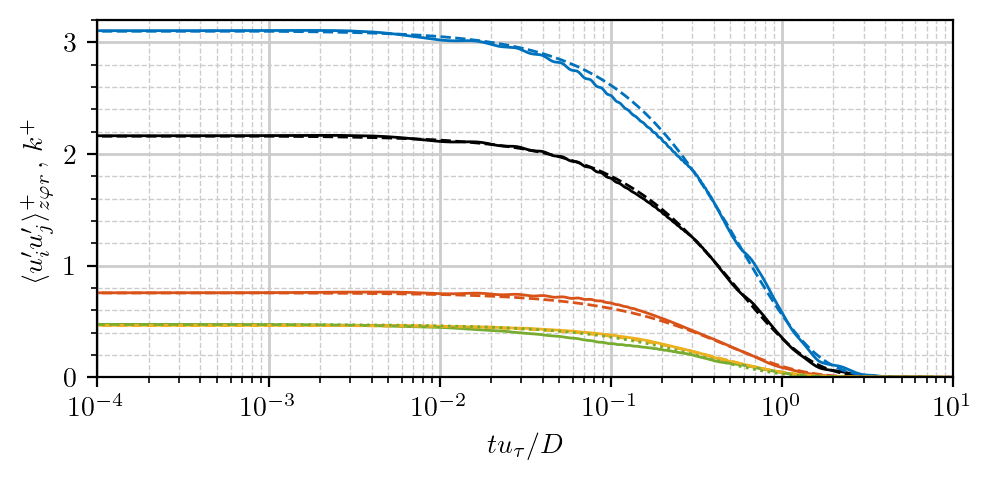}
\caption{
The decay of the Reynolds stresses $\langle u^\prime_iu^\prime_j\rangle^+_{z\varphi r}$ and the turbulent kinetic energy $k^+$ for case 1. 
\blackline, $k^+$; 
\blueline, $\langle u^\prime_zu^\prime_z\rangle^+_{z\varphi r}$; 
\redline, $\langle u^\prime_\varphi u^\prime_\varphi\rangle^+_{z\varphi r}$; 
\yellowline, $\langle u^\prime_r u^\prime_r\rangle^+_{z\varphi r}$; 
\greenline, $\langle u^\prime_r u^\prime_z\rangle^+_{z\varphi r}$.
Fitted exponential functions $\phi=\phi(t=0)\exp(-t/\tau)$:
\blackdashed, $\phi=k^+$, $\tau=0.55D/u_\tau$; 
\bluedashed, $\phi=\langle u^\prime_z u^\prime_z\rangle^+_{z\varphi r}$, $\tau=0.58D/u_\tau$; 
\reddashed, $\phi=\langle u^\prime_\varphi u^\prime_\varphi\rangle^+_{z\varphi r}$, $\tau=0.49D/u_\tau$;
\yellowdashed, $\phi=\langle u^\prime_r u^\prime_r\rangle^+_{z\varphi r}$, $\tau=0.42D/u_\tau$;
\greendotted, $\phi=\langle u^\prime_r u^\prime_z\rangle^+_{z\varphi r}$, $\tau=0.36D/u_\tau$
.
}
\label{fig:uiuj_vs_t}
\end{figure}

Besides to the bulk flow rate $u_b$, the streamwise traveling wave control influences the turbulent kinetic energy (TKE).
The time series of the volume-averaged TKE
\begin{equation}
k^+=\frac{1}{2}\left(\left\langle u^\prime_zu^\prime_z \right\rangle^+_{z \varphi r}
+\left\langle u^\prime_\varphi u^\prime_\varphi \right\rangle^+_{z \varphi r}
+\left\langle u^\prime_ru ^\prime_r \right\rangle^+_{z \varphi r}\right),
\label{eq:tke}
\end{equation}
obtained for the simulations listed in Table~\ref{tab:selcases} are presented in Fig.~\ref{fig:tke_vs_t}.
For UTWs (case 3) the blue dashed-dotted line in Fig.~\ref{fig:tke_vs_t} indicates a sudden TKE surge after the control is activated. This is followed by a decrease and another increase until the TKE settles after $t u_\tau/D \approx 7$ at a statistically stationary value of approximately $2.2$, which matches the value of the corresponding uncontrolled flow.
Bear in mind that although the flow remains turbulent, case 3 is accompanied by a drastic increase in the bulk flow rate and a decrease in drag (see Fig.~\ref{fig:rd_vs_t}) caused by the pumping effect~\cite{Hoepffner2009}.
For the standing wave in case 2 (dashed blue line), the TKE falls to $0.7$ at $tu_\tau D=0.6$, before rising again and settling at a value of approximately $1.3$. This corresponds to about 60\% of the TKE of the uncontrolled flow.
For case 4 (DTWs with $a^+=4.5$), the TKE fluctuates close to the level of the uncontrolled flow, after reaching a peak of nearly $4$ at $t u_\tau/D=2$.
Hightening the wave amplitude to $a^+=6$ relaminarizes case 1, lowering drag (blue line in Fig.~\ref{fig:rd_vs_t}) and TKE (blue line in Fig.~\ref{fig:tke_vs_t}).
Similar to the selected cases of relaminarization at a higher Reynolds number (case 5,6, and 7), the TKE of case 1 rapidly decays to zero.
The temporal evolution of the box-averaged TKE and Reynolds stresses $\langle u^\prime_iu^\prime_j\rangle^+_{z \varphi r}$ for case 1 is presented in Fig.~\ref{fig:uiuj_vs_t}. 
As indicated by the dashed lines in Fig.~\ref{fig:uiuj_vs_t}, the decay of the TKE and the Reynolds stresses can be modeled using an exponential decay law,
\begin{equation}
\phi(t)=\phi(t=0)\exp\left(-\frac{t}{\tau}\right),
\end{equation}
with the following values: $\tau=0.55D/u_\tau$ for $\phi=k^+$, $\tau=0.58D/u_\tau$ for $\phi=\langle u^\prime_z u^\prime_z\rangle^+_{z\varphi r}$, $\tau=0.49D/u_\tau$ for $\phi=\langle u^\prime_\varphi u^\prime_\varphi\rangle^+_{z\varphi r}$, $\tau=0.42D/u_\tau$ for $\phi=\langle u^\prime_r u^\prime_r\rangle^+_{z\varphi r}$, and $\tau=0.36D/u_\tau$ for $\phi=\langle u^\prime_r u^\prime_z\rangle^+_{z\varphi r}$.
Even though 99\% of the TKE and the Reynolds shear stress are eliminated in less than three dimensionless time units in the relaminarizaion cases, accelerating the bulk flow to its terminal velocity takes much longer (see Fig~\ref{fig:rd_vs_t}).
When it comes to relaminarization, however, the temporal evolution of the TKE, the Reynolds stresses, and the drag reduction rate scale when the time step is normalized by $D/u_\tau$, independent of the Reynolds number.

\subsection{Upstream traveling waves ($c<0$)}
\label{ss:utw}
\begin{figure} 
\centering
\includegraphics[]{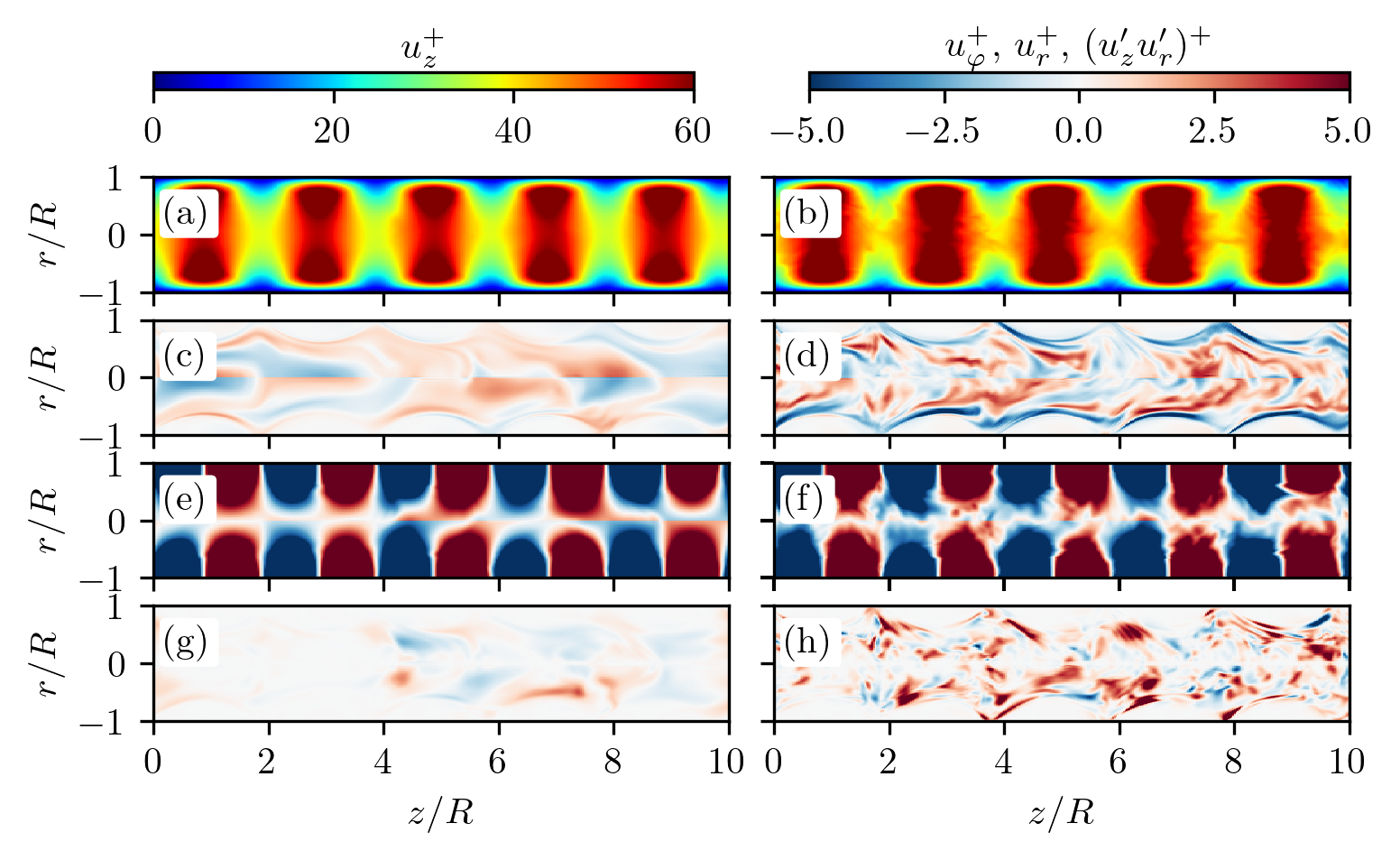}
\caption{
Instantaneous flow field realization of UTW controlled pipe flow (case 3) at $t u_\tau/D=4.0$ (a,c,e,g) and $t u_\tau/D=43.3$ (b,d,f,h) after the start of the control. Velocity components $u_z^+$ (a,b) $u^+_\varphi$ (c,d), and $u^+_r$ (e,f). Instantaneous Reynolds shear stress $(u_z^\prime u_r^\prime)^+$ (g,h). 
Videos of volume-rendered flow quantities can be found in the supplementary material.
}
\label{fig:inst_utw}
\end{figure}

Fig.~\ref{fig:inst_utw} portrays the instantaneous flow field realizations of the three velocity components and the Reynolds shear stress in an $rz$-plane for UTWs (case 3) at $t=4.0D/u_\tau$ (left column) and at $t=43.3D/u_\tau$ after the flow has reached a statistically stationary state (right column).
The streamwise velocity component $u^+_z$ exhibits periodic acceleration and deceleration (Fig.~\ref{fig:inst_utw}a,b), which correspond to the periodic wall-normal blowing and suction visible in the radial velocity component (Fig.~\ref{fig:inst_utw}e,f).
The instantaneous streamwise and radial velocities are dominated by the mean flow.
Nevertheless, the azimuthal velocity component (Fig.~\ref{fig:inst_utw}c,d) consists of turbulent structures characteristic of UTW-controlled turbulent pipe flow.
These structures seem to be suppressed by the control during the transient phase (Fig.~\ref{fig:inst_utw}c), but they grow more intense at $t=43.3D/u_\tau$ (Fig.~\ref{fig:inst_utw}d).
Compared to the turbulence structures in uncontrolled pipe flow, these turbulent fluctuations in UTW-controlled pipe flow are pushed away from the wall owing to the suction and blowing (Fig.~\ref{fig:inst_utw}c,d).
This effect is also visible in the instantaneous Reynolds shear stress, as shown in Fig.~\ref{fig:inst_utw}(g,h). 
Overall, the drag increase due to the mostly positive turbulent Reynolds shear stress structures is overcompensated by the acceleration of the flow due to the pumping effect for UTWs, leading to a net drag reduction (cf. Fig.~\ref{fig:rd_vs_t}).
\subsection{Downstream traveling waves ($c>0$)}
\label{ss:dtw}
\begin{figure} 
\centering
\includegraphics[]{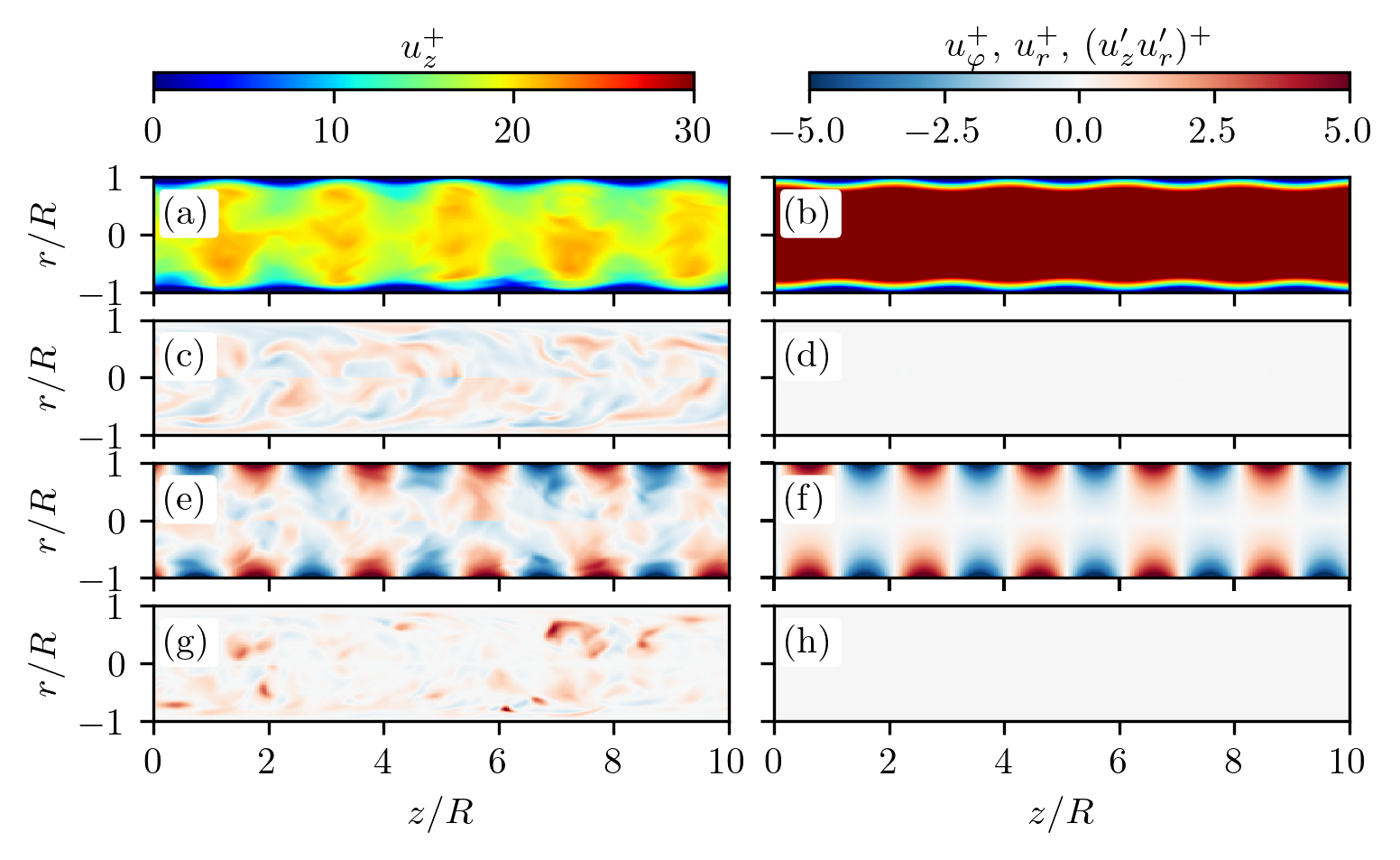}
\caption{
Instantaneous flow field realization of case 1 ($a^+=6$, $\lambda^+=360$, $c^+=90$) at $t u_\tau/D=0.5$ (a,c,e,g) $t u_\tau/D=114.3$ (b,d,f,h) after the start of the control. Velocity components $u_z^+$ (a,b) $u^+_\varphi$ (c,d), and $u^+_r$ (e,f). Instantaneous Reynolds shear stress $(u_z^\prime u_r^\prime)^+$ (g,h). 
Videos of volume-rendered flow quantities can be found in the supplementary material.
}
\label{fig:inst_case1}
\end{figure}

\begin{figure} 
\centering
\includegraphics[trim=8 0 8 0,clip]{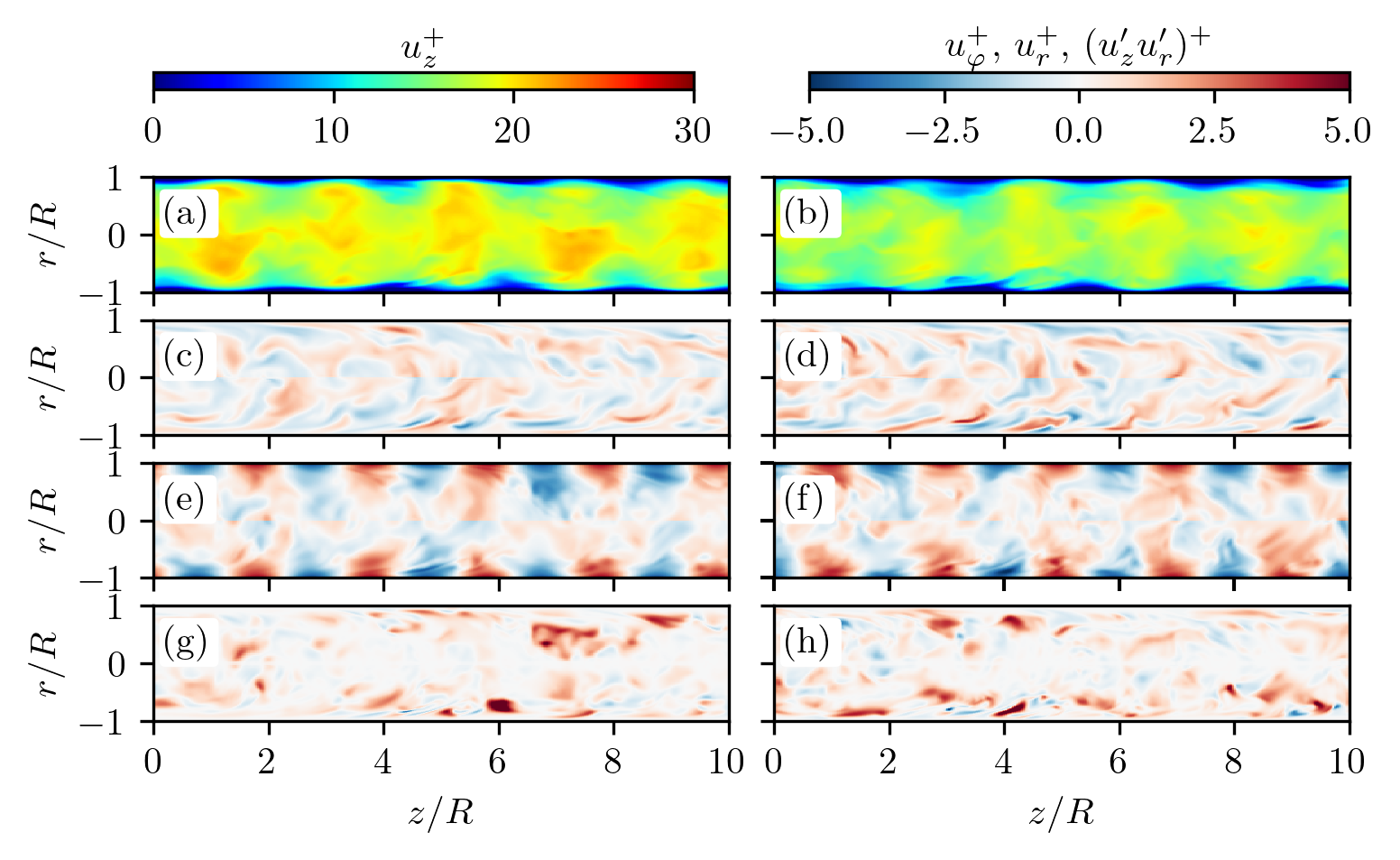}
\caption{
Instantaneous flow field realization of case 4 ($a^+=4.5$, $\lambda^+=360$, $c^+=90$) at $t u_\tau/D=0.5$ (a,c,e,g) and $t u_\tau/D=114.3$ (b,d,f,h) after the start of the control. Velocity components $u_z^+$ (a,b) $u^+_\varphi$ (c,d), and $u^+_r$ (e,f). Instantaneous Reynolds shear stress $(u_z^\prime u_r^\prime)^+$ (g,h). 
}
\label{fig:inst_case4}
\end{figure}

For the DTWs, the instantaneous flow field realizations of the velocity components and the Reynolds shear stress at times $t=0.5D/u_\tau$ and $t=114.3D/u_\tau$  are presented in Fig.~\ref{fig:inst_case1} and Fig.~\ref{fig:inst_case4}, respectively, for case~1 and case~4. Case~1  relaminarizes, while case~4 does not.
In case~1, the DTWs effectively dampen turbulent fluctuations near the wall. This is reflected by the decay to zero of the azimuthal velocity component (Fig.~\ref{fig:inst_case1}c,d) and the Reynolds shear stress  (Fig.~\ref{fig:inst_case1}g,h).
Additionally, the initially turbulent streamwise velocity component in Fig.~\ref{fig:inst_case1}(a) accelerates because of the transition to a laminar velocity profile lacking turbulent fluctuations in Fig.~\ref{fig:inst_case1}(b).
For case~4, the traveling wave amplitude of $a^+=4.5$ and the effective layer width of $\eta^+=2.9$ are too small to dampen the near-wall turbulence. This is manifested in the lack of significant changes in the azimuthal and radial velocity components and the Reynolds shear stress between $t u_\tau/D=0.5$ (Fig.~\ref{fig:inst_case4}c,e,g) and $t u_\tau/D=114.3$ (Fig.~\ref{fig:inst_case4}d,f,h).
Nevertheless, the effective blocking of the pipe cross-section due to the induced traveling waves leads to a deceleration of the streamwise velocity component, as seen in Fig.~\ref{fig:inst_case4}(a,b), and thus, to a reduction in bulk flow rate and an increase in drag compared to the uncontrolled flow (see Fig.~\ref{fig:rd_vs_t}).\par
%
%
%
\subsection{Time- and phase-averaged statistics}
\label{ss:timesaverage}
\begin{figure} 
\centering
\includegraphics[]{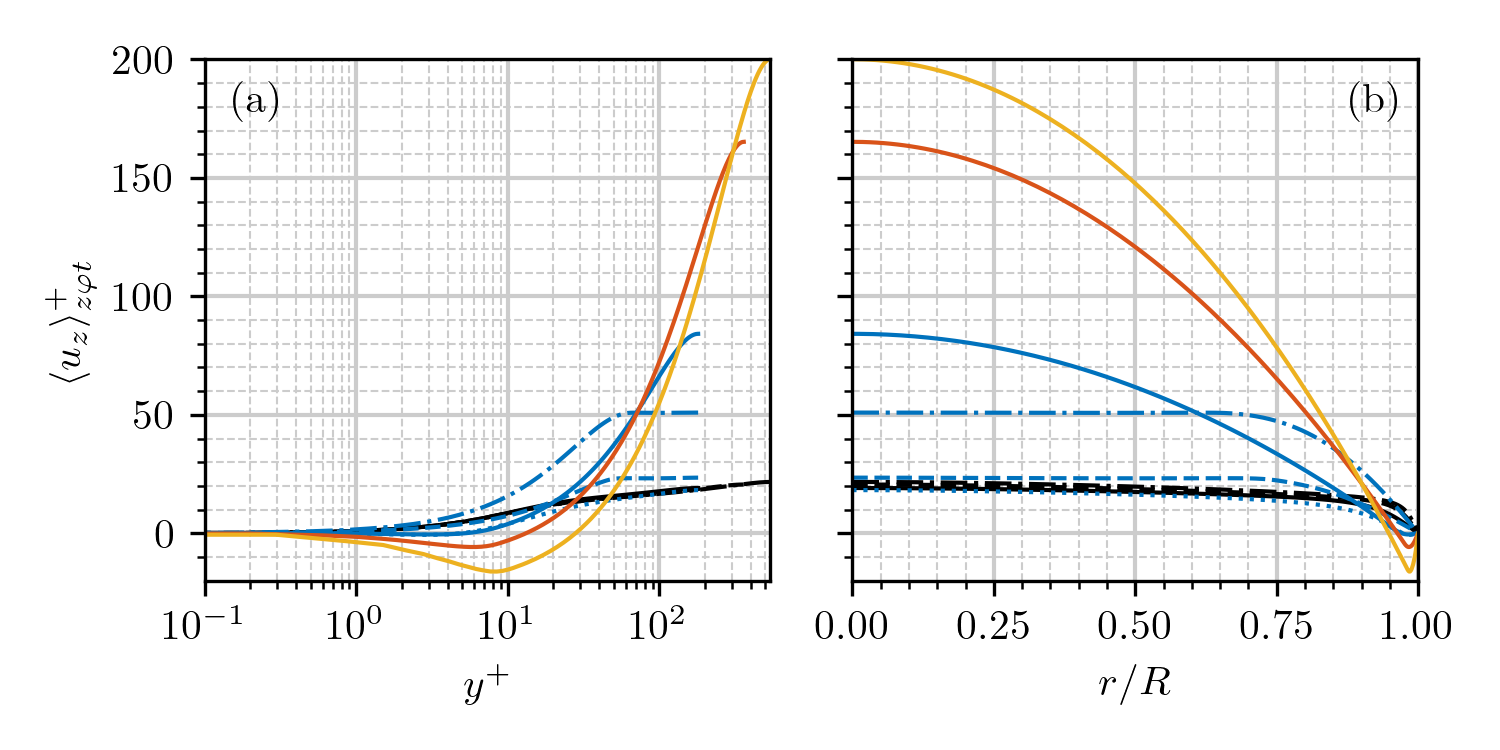}
\caption{
Mean streamwise velocity profile $\langle u_z \rangle^+_{z\varphi t}$ for selected flow cases: \blueline, case 1; \bluedashed, case 2; \bluedashdotted, case 3; \bluedotted, case 4; \redline, case 5; \yellowline, case 6; \purpleline, case 7; see Table~\ref{tab:selcases}. Black lines indicate reference cases of uncontrolled flow: \blackline, $\mathrm{Re}_\tau=180$~\cite{Bauer2017}, \blackdashed, $\mathrm{Re}_\tau=360$~\cite{Bauer2017}, \blackdashdotted, $\mathrm{Re}_\tau=540$.
}
\label{fig:umean}
\end{figure}

\begin{figure} 
\centering
\includegraphics[]{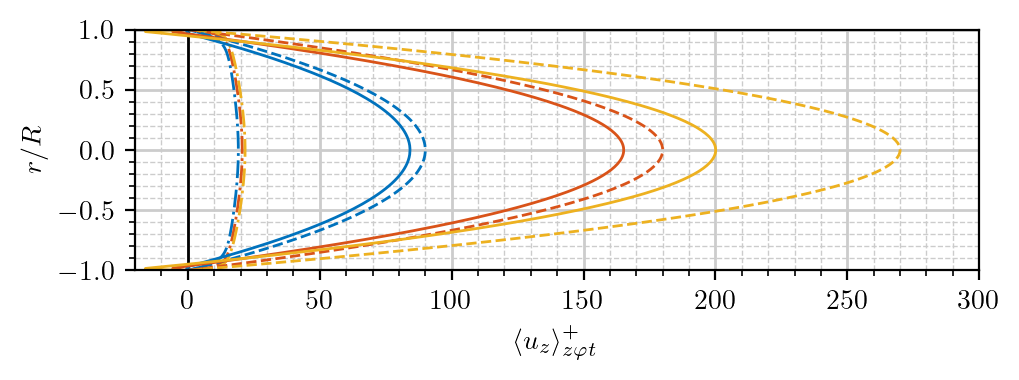}
\caption{
Mean streamwise velocity profile $\langle u_z \rangle^+_{z\varphi t}$ for relaminarization cases: \blueline, case 1 ($\mathrm{Re}_\tau=180$); \redline, case 5 ($\mathrm{Re}_\tau=360$); \yellowline, case 6 ($\mathrm{Re}_\tau=540$); \purpleline, case 7 ($\mathrm{Re}_\tau=720$); see Table~\ref{tab:selcases}. Dashed lines indicate laminar Hagen-Poiseuille flow profiles at corresponding friction Reynolds numbers. Dashed-dotted lines indicate reference cases of uncontrolled turbulent flow at corresponding friction Reynolds numbers~\cite{Bauer2017}.
}
\label{fig:umlam}
\end{figure}

\begin{figure} 
\centering
\includegraphics[]{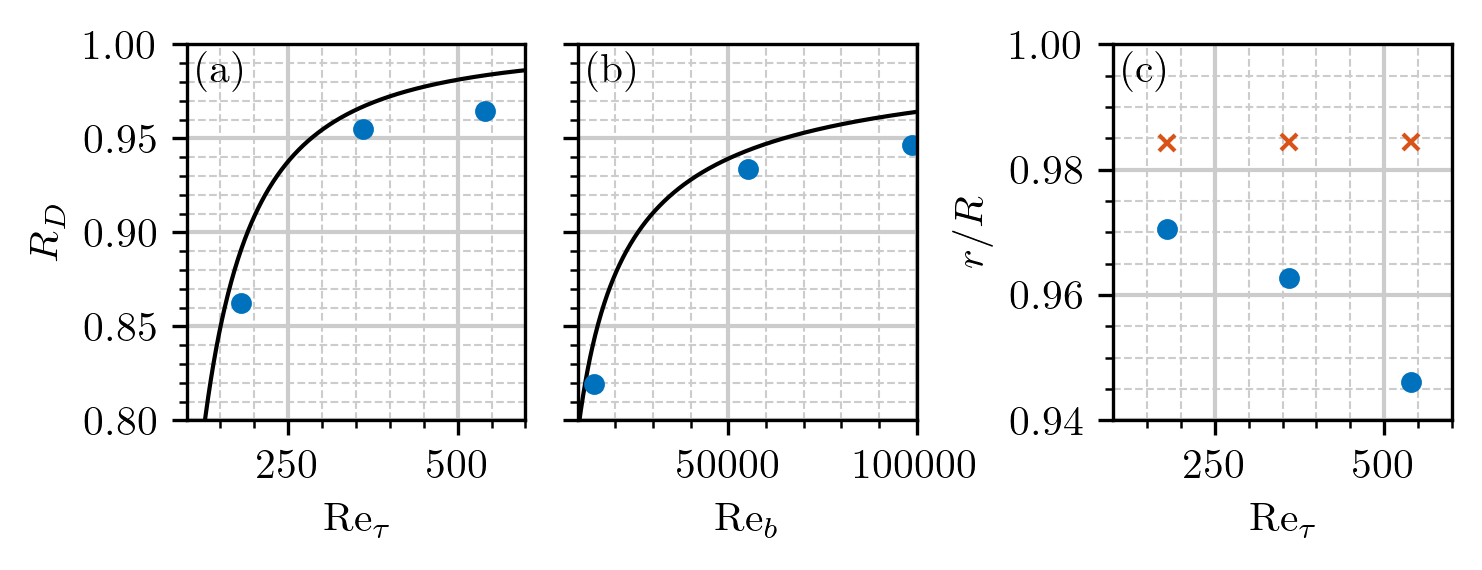}
\caption{
Drag reduction rate $R_D$ versus the friction Reynolds number $\mathrm{Re}_\tau$ (a) and the bulk Reynolds number $\mathrm{Re}_\tau$ (b), as well as the radius where the mean velocity profile has its minimum (red crosses) and the radius $r_0/R$, where $\langle u_z \rangle_{z\varphi t}=0$ (blue dots) versus the friction Reynolds number $\mathrm{Re}_\tau$ for relaminarization cases. Lines indicate the theoretical limit of the corresponding laminar flow at constant $\mathrm{Re}_\tau$ (a) and $\mathrm{Re}_\tau$ (b).
}
\label{fig:umradius}
\end{figure}

After reaching the statistically stationary state shown in Fig.~\ref{fig:rd_vs_t}, the mean velocity profile is computed by averaging over time, as well as over $z$ and $\varphi$.
Fig.~\ref{fig:umean} displays the mean streamwise velocity profile $\langle u_z\rangle^+_{z \varphi t}$, which is normalized in wall units and plotted logarithmically against the distance from the pipe wall in wall units $y^+$ in Fig.~\ref{fig:umean}(a) and plotted linearly against the pipe radius $r/R$ in Fig.~\ref{fig:umean}(b).
As indicated by Fig.~\ref{fig:umean}(a), all selected cases except case~2 deviate significantly from the logarithmic behavior of the corresponding uncontrolled flow (indicated by black lines).
The mean streamwise velocity profiles of the relaminarization cases (solid colored lines) show deceleration near the wall. This  leads to negative mean velocity minima at $y^+\approx2.8$ for case 1, $y^+\approx5.6$ for case 5, and $y^+\approx8.6$ for case 6. 
As demonstrated in Fig.~\ref{fig:umean}(b), these minima collapse at $r/R\approx0.98$ when the radial coordinate is normalized by the pipe radius.
In the bulk flow region, unlike near the wall, the relaminarization of cases 1,5, and 6 prompts a strong acceleration of the flow and the characteristic parabolic laminar flow profile (see Fig.~\ref{fig:umean}b).
For case 4 (DTW control without relaminarization), the flow is slightly decelerated, particularly near the pipe wall.
For UTWs (case 3), the flow accelerates both near the wall (Fig.~\ref{fig:umean}a) and away from the wall (Fig.~\ref{fig:umean}b), where the mean velocity profile features a plateau ($0 \le r/R \le0.75$).
In the case of a standing wave (case 2), the mean velocity profile only slightly deviates from the profile of the corresponding uncontrolled flow for $y^+<20$ (see  Fig.~\ref{fig:umean}a). In the bulk flow region, however, the flow accelerates and forms a plateau-shaped velocity profile with approximately half the level of the UTW profile.
Fig.~\ref{fig:umlam} compares the mean velocity profiles of the relaminarization cases (solid lines) to the turbulent mean velocity profiles of the corresponding uncontrolled flow (dashed-dotted lines) and the theoretical laminar Hagen-Poiseuille flow profiles (dashed lines).
Remarkably, unlike the theoretical laminar flow profiles, the mean velocity profiles obtained from the relaminarization cases 1, 5, and 6 feature a backflow region close to the wall, induced by the DTW boundary condition (see also Fig.~\ref{fig:umean}).
Hence, the effective diameter of the controlled pipe flow decreases compared to the theoretical laminar flow at the same friction Reynolds number, which, in turn, leads to a deviation of the mean flow profile from the theoretical laminar profile in the bulk region, as seen in Fig.~\ref{fig:umlam}.
The scaling of the radius of the mean velocity minimum and $r_0$, defined as $\langle u\rangle_{z\varphi t}(r=r_0)=0$, is presented in Fig.~\ref{fig:umradius}, together with the scaling of the drag reduction rate.
In Fig.~\ref{fig:umradius}, the drag reduction rate is computed using expression (\ref{eq:rd}), where $C_{f0}$ is evaluated at the corresponding friction Reynolds number (a) and the corresponding bulk Reynolds number using expression (\ref{eq:ubfit}) (b), respectively, generating slightly different results.
Thus, the deviation of the DTW-controlled laminar velocity profiles from the theoretical laminar profiles in Fig.~\ref{fig:umlam} is reflected in the offset between the achieved drag reduction and the theoretical limit in Fig.~\ref{fig:umradius}(a). 
Notwithstanding, the present flow control approach achieves 97\% or more of the theoretically accessible drag reduction.\par

\begin{figure} 
\centering
\includegraphics[]{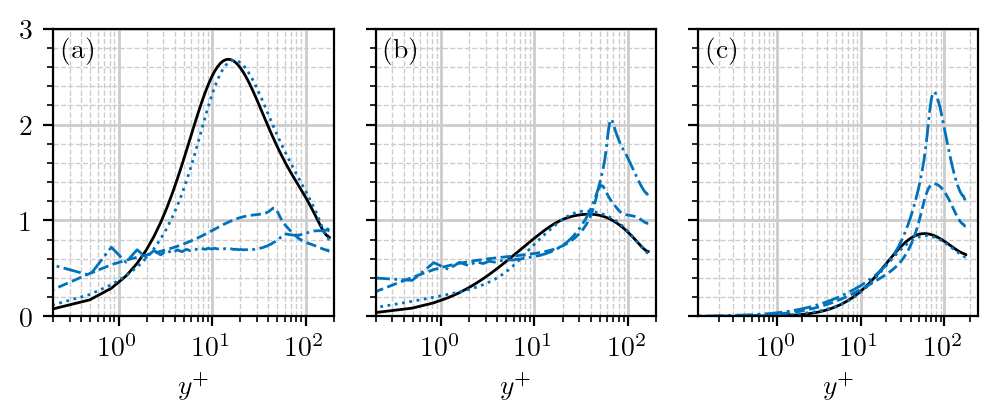}
\caption{
Turbulent intensities $u_{z,rms}^+$ (a), $u_{z,rms}^+$ (b), and $u_{z,rms}^+$ (c) for selected flow cases: \blueline, case 1; \bluedashed, case 2; \bluedashdotted, case 3; \bluedotted, case 4, see Table~\ref{tab:selcases}. Black lines indicate the reference cases of uncontrolled turbulent pie flow at $\mathrm{Re}_\tau=180$~\cite{Bauer2017}.
}
\label{fig:urms}
\end{figure}

Similar to the mean velocity profiles, the turbulence intensity profiles are computed by applying temporal averaging, as well as azimuthal and axial averaging, to the normal turbulent Reynolds components before taking the square root, $u^+_{i,rms}=\sqrt{\langle u_i^{\prime\prime} u_i^{\prime\prime}\rangle^+_{z \varphi t}}$.
Fig.~\ref{fig:urms} shows the turbulence intensities for cases~1--4.
Pertaining to DTWs without relaminarization (case~4), the turbulence intensity profiles exhibit minimal deviation from the corresponding uncontrolled case (solid black lines).
Conversely, the turbulence intensities for the case with relaminarization (case~1) are zero, as expected.
A similar behavior can be observed in the cases of the standing wave (case 2) and UTW (case~3),
where energy is transferred from the streamwise component (Fig.~\ref{fig:urms}a) to the azimuthal and radial components (Fig.~\ref{fig:urms}b,c).
Consequently, the characteristic $u^+_{z,rms}$ peak at $y^+\approx15$ vanishes, and new peaks emerge in $u^+_{\varphi,rms}$ and $u^+_{r,rms}$ around $50 \lesssim   y^+ \lesssim  80$.
Accordingly, the traveling wave flow control with $c \le 0$ effectively destroys the characteristic streaky structure of the buffer layer with its dominant $u_z$-modes.
Therefore, energy is transferred to the cross-sectional velocity fluctuations at a greater distance from the wall.\par
\begin{figure} 
\centering
\includegraphics[]{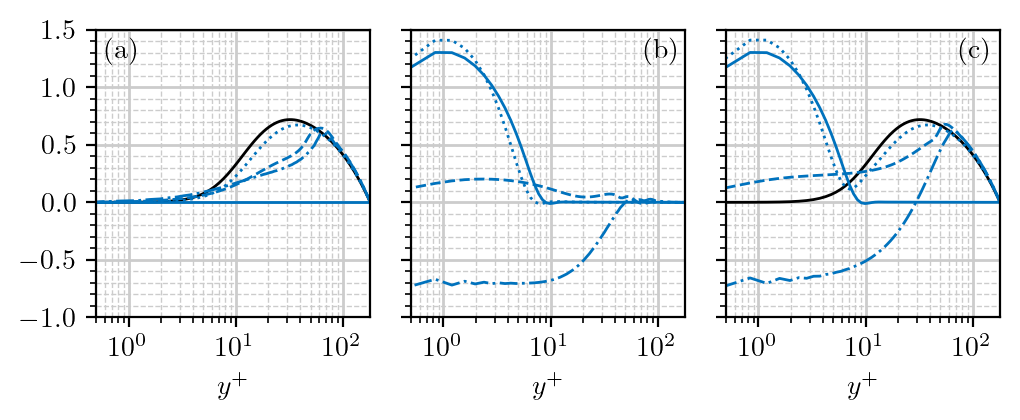} 
\caption{
Reynold shear stresses: $\langle u^{\prime\prime}_z u^{\prime\prime}_r\rangle_{z\varphi t}$, (a); $\langle \tilde{u}_z \tilde{u}_r\rangle_{z\varphi t}$ (b); $\langle u^{\prime\prime}_z u^{\prime\prime}_r\rangle_{z\varphi t}$+$\langle \tilde{u}_z \tilde{u}_r\rangle_{z\varphi t}$, (c), for selected flow cases: \blueline, case 1; \bluedashed, case 2; \bluedashdotted, case 3; \bluedotted, case 4, see Table~\ref{tab:selcases}. Black lines indicate the reference cases of uncontrolled turbulent pie flow at $\mathrm{Re}_\tau=180$~\cite{Bauer2017}.
}
\label{fig:rss}
\end{figure}

For the cases~1 to~4, Fig.~\ref{fig:rss} illustrates the spatio-temporal average of the turbulent Reynolds shear stress $\langle u^{\prime\prime}_z u^{\prime\prime}_r\rangle_{z\varphi t}$ (a), the periodic Reynolds shear stress $\langle \tilde{u}_z \tilde{u}_r\rangle_{z\varphi t}$ (b), as well as the total Reynolds shear stress (c).
For case~1 (solid line), the DTWs dampen the turbulent Reynolds shear stress to zero (Fig.~\ref{fig:rss}a), while inducing a positive periodic Reynolds shear stress close to the wall (Fig.~\ref{fig:rss}b).
The latter leads to an increase in drag in comparison to the Hagen-Poiseuille flow (cf. Fig.~\ref{fig:umlam}).
For case~2 (dashed line), the turbulent Reynolds shear stress is reduced compared to the uncontrolled flow (black line, Fig.~\ref{fig:rss}a), while the periodic Reynolds shear is moderately positive (Fig.~\ref{fig:rss}b).
In total, given the restriction of the periodic Reynolds shear stress to the proximitiy of the wall, the $y$-integrated overall shear stress is reduced relative to the uncontrolled case (black line, Fig.~\ref{fig:rss}c), which results in a drag reduction for case~2.
For case~3 (dashed-dotted line), both the turbulent Reynolds shear stress (Fig.~\ref{fig:rss}a) and the periodic Reynolds shear stress (Fig.~\ref{fig:rss}b) are lessened in comparison to the uncontrolled flow.
The latter phenomenon is attributed to the pumping effect, which induces a negative Reynolds shear stress near the wall.
This, in turn, results in a substantial drag reduction for case~3.
For case~4 (dotted line), the turbulent Reynolds shear stress is slightly decreased in comparison to the uncontrolled flow (Fig.~\ref{fig:rss}a),
whereas the periodic Reynolds shear is similar to case~1 (Fig.~\ref{fig:rss}b).
In total, the Reynolds shear stress is greater than in the uncontrolled case (Fig.~\ref{fig:rss}c), causing an increase in drag for case~4.
\par
\begin{figure} 
\centering
\includegraphics[]{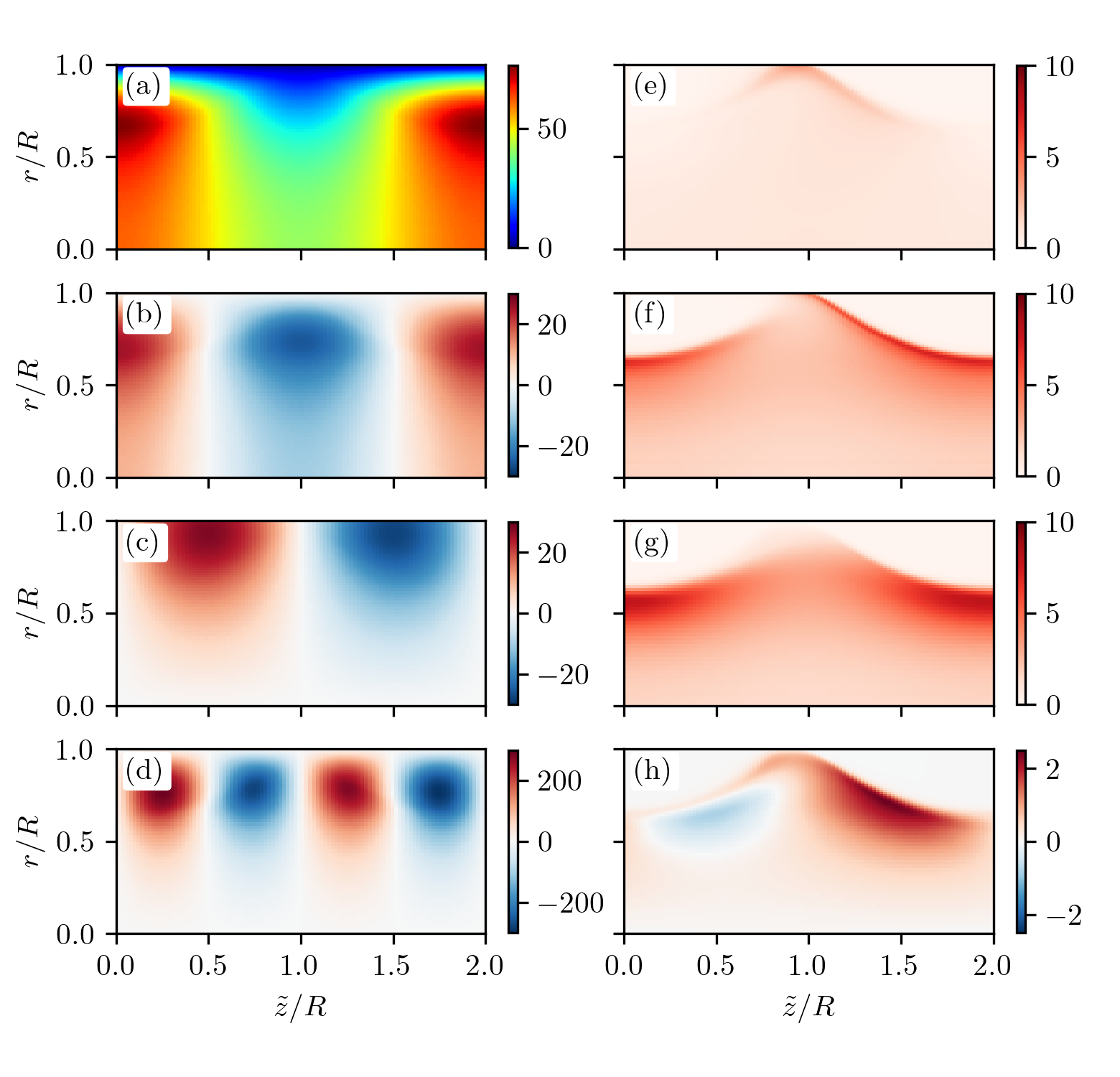}
\caption{Phase-averaged statistics for case 3 ($a^+=27$, $\lambda^+=360$, $c^+=-9$). 
(a) Phase-averaged streamwise velocity $\langle u_z\rangle_{N_{\phi_z}\varphi t}$; 
(b) periodic component of the streamwise velocity $\tilde{u}_z$; 
(c) phase-averaged radial velocity $\langle u_r\rangle_{N_{\phi_z}\varphi t}=\tilde{u}_r$; 
(d) periodic Reynolds shear stress $\tilde{u}_z \tilde{u}_r$;
(e) turbulent streamwise Reynolds stress $\langle u_z^{\prime\prime} u_z^{\prime\prime} \rangle_{N_{\phi_z \varphi t}}$
(f) turbulent azimuthal Reynolds stress $\langle u_\varphi^{\prime\prime} u_\varphi^{\prime\prime} \rangle_{N_{\phi_z \varphi t}}$
(g) turbulent radial Reynolds stress $\langle u_r^{\prime\prime} u_r^{\prime\prime} \rangle_{N_{\phi_z \varphi t}}$
(h) turbulent Reynolds shear stress $\langle u_z^{\prime\prime} u_r^{\prime\prime} \rangle_{N_{\phi_z \varphi t}}$.
}
\label{fig:phase3}
\end{figure}
\begin{figure} 
\centering
\includegraphics[]{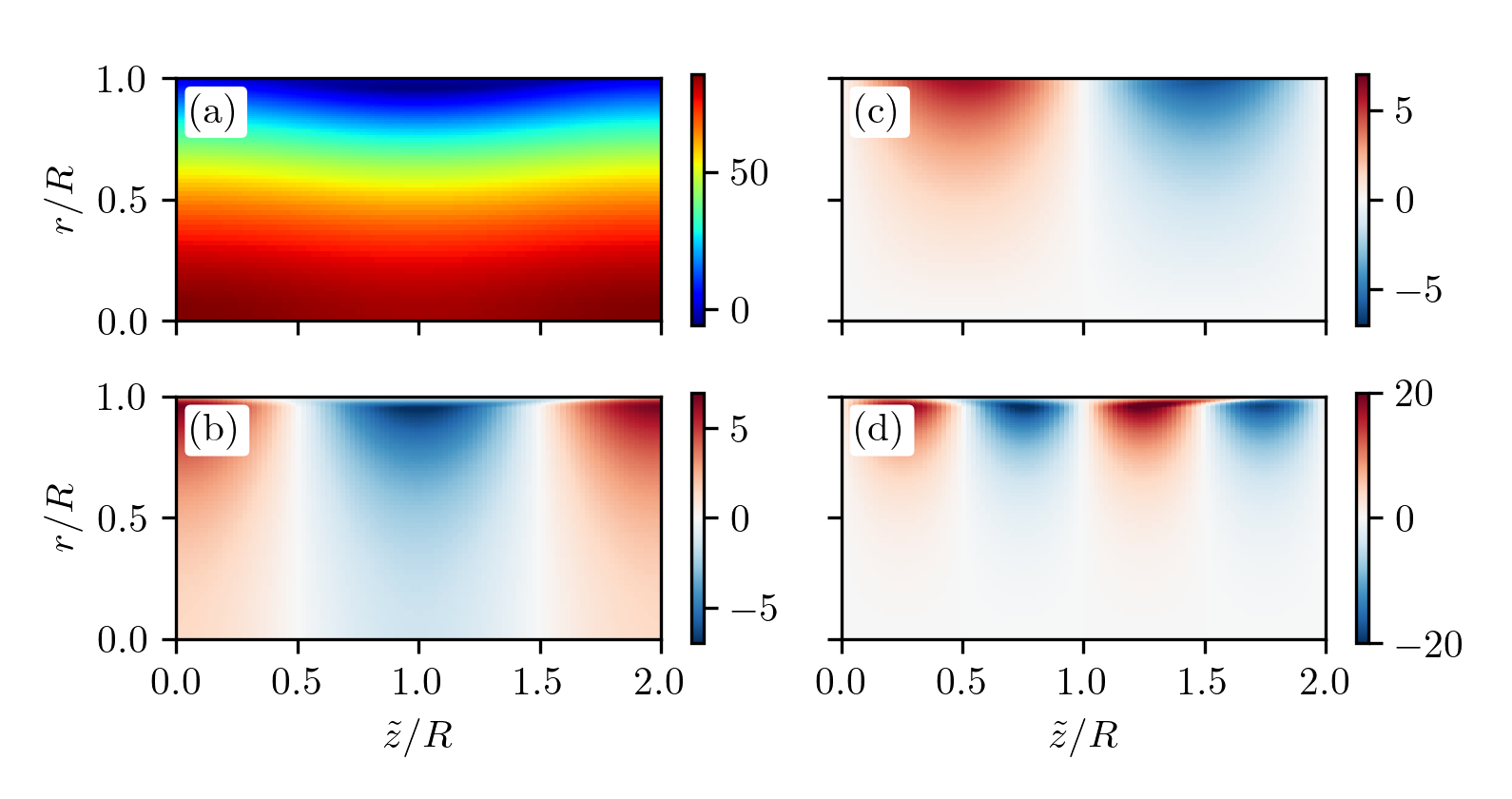}
\caption{Phase-averaged statistics for case 1 ($a^+=6$, $\lambda^+=360$, $c^+=90$). 
(a) Phase-averaged streamwise velocity $\langle u_z\rangle_{N_{\phi_z}\varphi t}$; 
(b) periodic component of the streamwise velocity $\tilde{u}_z$; 
(c) phase-averaged radial velocity $\langle u_r\rangle_{N_{\phi_z}\varphi t}=\tilde{u}_r$; 
(d) periodic Reynolds shear stress $\tilde{u}_z \tilde{u}_r$;
}
\label{fig:phase1}
\end{figure}
\begin{figure} 
\centering
\includegraphics[]{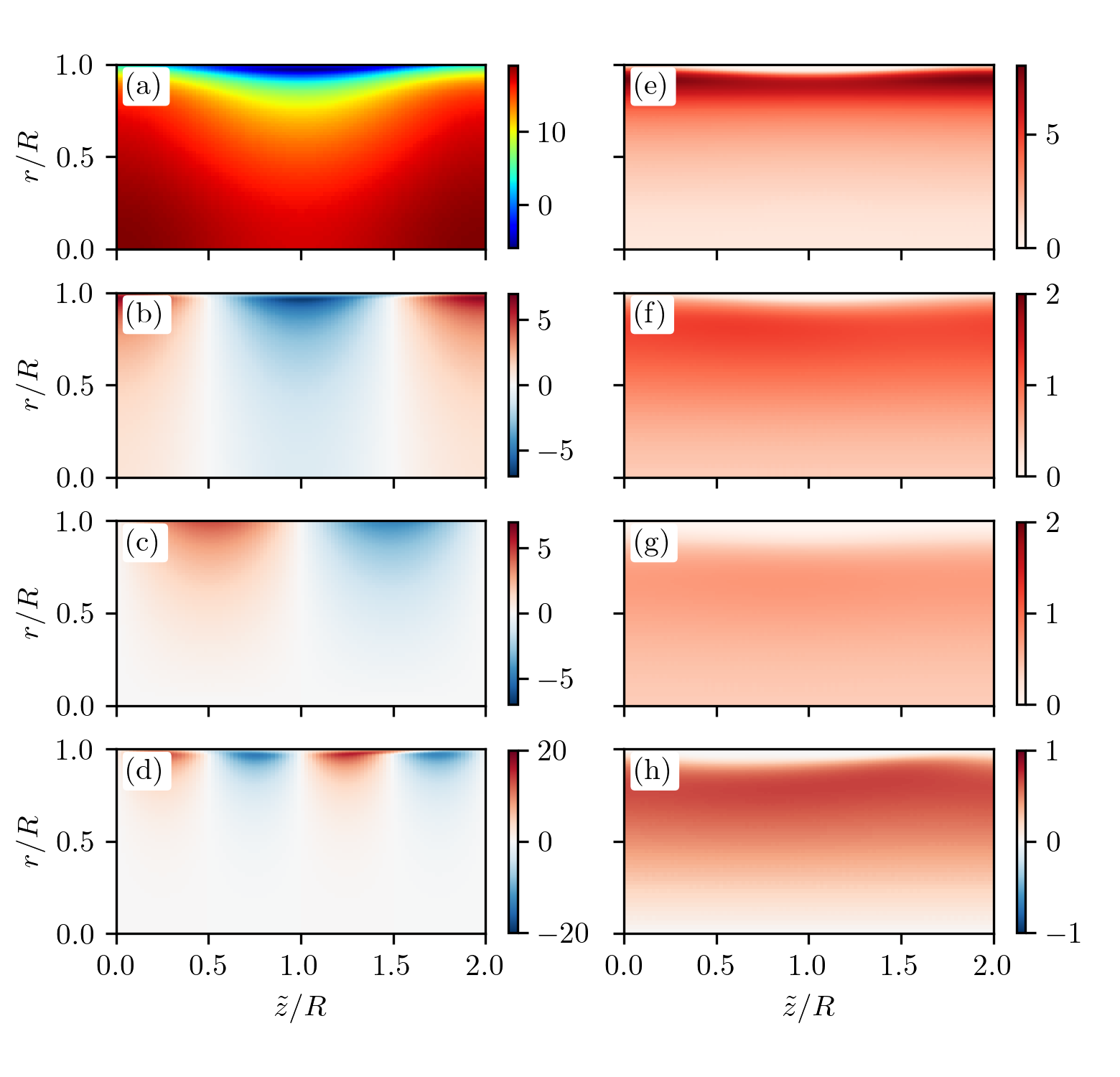}
\caption{Phase-averaged statistics for case 4 ($a^+=4.5$, $\lambda^+=360$, $c^+=90$). 
(a) Phase-averaged streamwise velocity $\langle u_z\rangle_{N_{\phi_z}\varphi t}$; 
(b) periodic component of the streamwise velocity $\tilde{u}_z$; 
(c) phase-averaged radial velocity $\langle u_r\rangle_{N_{\phi_z}\varphi t}=\tilde{u}_r$; 
(d) periodic Reynolds shear stress $\tilde{u}_z \tilde{u}_r$;
(e) turbulent streamwise Reynolds stress $\langle u_z^{\prime\prime} u_z^{\prime\prime} \rangle_{N_{\phi_z \varphi t}}$
(f) turbulent azimuthal Reynolds stress $\langle u_\varphi^{\prime\prime} u_\varphi^{\prime\prime} \rangle_{N_{\phi_z \varphi t}}$
(g) turbulent radial Reynolds stress $\langle u_r^{\prime\prime} u_r^{\prime\prime} \rangle_{N_{\phi_z \varphi t}}$
(h) turbulent Reynolds shear stress $\langle u_z^{\prime\prime} u_r^{\prime\prime} \rangle_{N_{\phi_z \varphi t}}$.
}
\label{fig:phase4}
\end{figure}
\begin{figure} 
\centering
\includegraphics[]{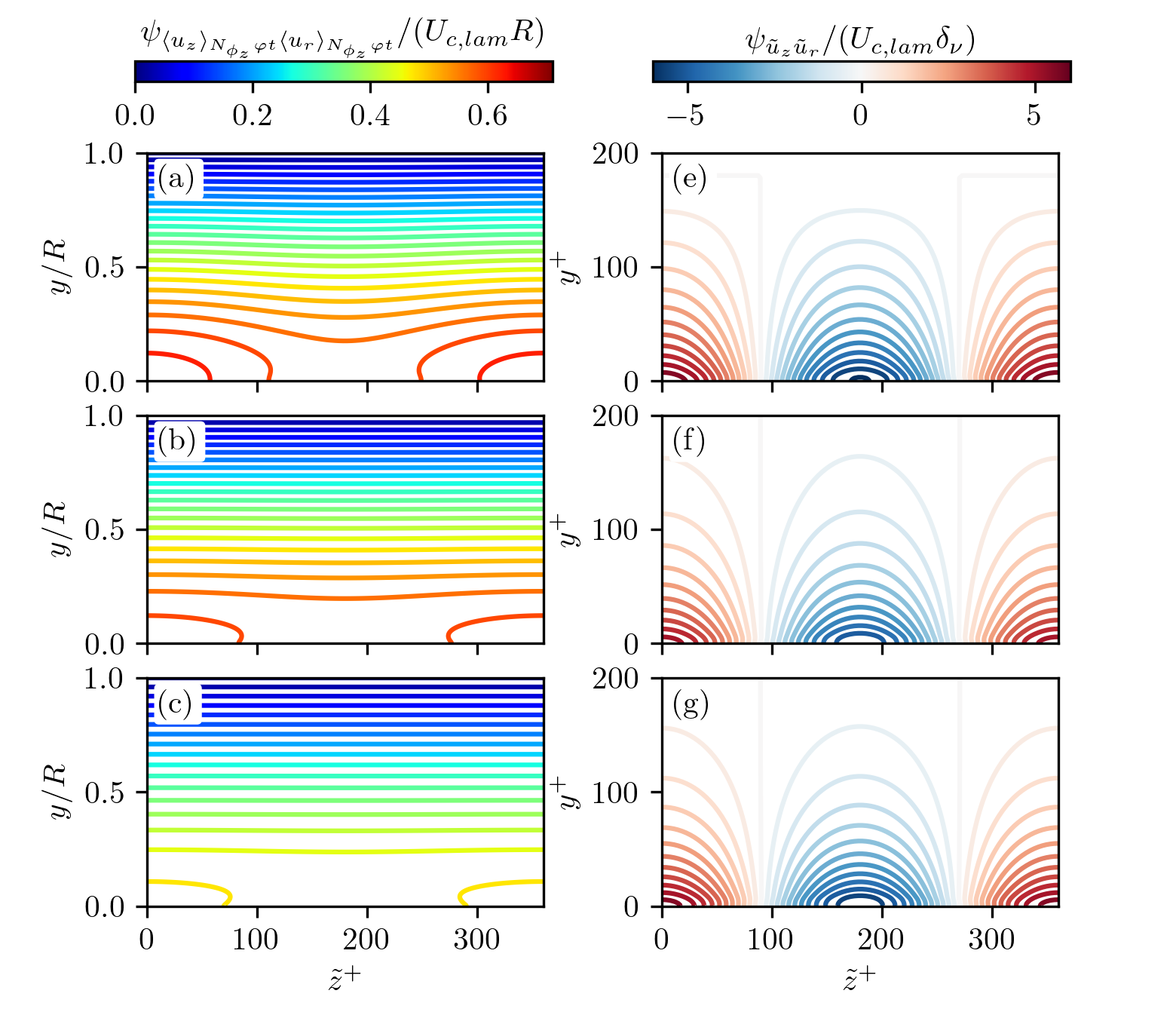}
\caption{stream function of the phase-averaged streamwise and radial velocity normalized by $U_{c,lam}$ and $R$ (a,b,c) and stream function of the periodic component of the streamwise and radial velocity normalized by $U_{c,lam}$ and $\delta_\nu$ (e,f,g) for case~1 (a,e), case~5 (b,f), and case~5 (c,g).
}
\label{fig:phasepsi}
\end{figure}

%
In the following, phase-averaged statistics evaluated using equation (\ref{eq:phaseavg}) are analyzed.
Firstly, the phase-averaged statistics for case~3 are displayed in Fig.~\ref{fig:phase3}.
As depicted in Fig.~\ref{fig:phase3}(a), the phase-averaged streamwise velocity component oscillates strongly for UTW-controlled flow.
The periodic component of the streamwise velocity $\tilde{u}_z$ is featured in Fig.~\ref{fig:phase3}(b).
This component is obtained by subtracting the $z\varphi t$-averaged streamwise velocity profile $\langle u_z\rangle_{z \varphi t}$ (cf. Fig.~\ref{fig:umean}) from the phase-averaged streamwise velocity (Fig.~\ref{fig:phase3}a).
Given that the $z\varphi t$-average of the radial velocity equals zero, the phase-averaged radial velocity, as presented in Fig.~\ref{fig:phase3}(c), corresponds to the periodic component of the radial velocity.
The product of the periodic streamwise and radial velocity components is the periodic Reynolds shear stress (Fig.~\ref{fig:phase3}d), the $\tilde{z}$-average of which is equivalent to the periodic Reynolds shear stress profile illustrated in Fig.~\ref{fig:rss}(b, dashed-dotted line).
The right column of Fig.~\ref{fig:phase3} shows the Reynolds stresses $\langle u_z^{\prime\prime} u_z^{\prime\prime} \rangle_{N_{\phi_z}\varphi t}$ (e), $\langle u_\varphi^{\prime\prime} u_\varphi^{\prime\prime} \rangle_{N_{\phi_z}\varphi t}$ (f), $\langle u_r^{\prime\prime} u_r^{\prime\prime} \rangle_{N_{\phi_z}\varphi t}$ (g), and $\langle u_z^{\prime\prime} u_r^{\prime\prime} \rangle_{N_{\phi_z}\varphi t}$ (h).
In contrast to the uncontrolled flow, the streamwise Reynolds stress (Fig.~\ref{fig:phase3}e) is less pronounced than the other two normal Reynolds stress components (Fig.~\ref{fig:phase3}f,g).
Furthermore, the region in which the azimuthal Reynolds stress (Fig.~\ref{fig:phase3}f) and the radial Reynolds stress (Fig.~\ref{fig:phase3}g) are maximal, $\tilde{z}/R=0$, $r/R\approx0.6$, corresponds to the region of maximum streamwise acceleration (Fig.~\ref{fig:phase3}a,b) and zero blowing/suction (Fig.~\ref{fig:phase3}c).
In contrast, the phase-averaged turbulent Reynolds shear stress (Fig.~\ref{fig:phase3}h) exhibits a positive maximum at $\tilde{z}/R\approx1.5$ and a weaker negative minimum at $\tilde{z}/R\approx0.5$. These streamwise coordinates align with the locations of maximum blowing and suction, respectively (cf. Fig.~\ref{fig:phase3}c).
Fig.~\ref{fig:phase1} presents the phase-averaged statistics for case~1, with the exclusion of turbulent quantities that are equal to zero due to relaminarization.
As illustrated in Fig.~\ref{fig:phase1}(a), the phase-averaged streamwise velocity displays a backflow region around $\tilde{z}=R$, which corresponds to the backflow region in the periodic component of the streamwise velocity, as depicted in Fig.~\ref{fig:phase1}(b). 
In comparison with case~3, the lower traveling wave amplitude results in diminished periodic components of the streamwise velocity (Fig.~\ref{fig:phase1}b) and the radial velocity (Fig.~\ref{fig:phase1}c), which are smaller in intensity and are confined to regions that are closer to the wall. 
This results in a periodic Reynolds shear stress that is one order of magnitude smaller in case~1 (Fig.~\ref{fig:phase1}d) than in case~3 (Fig.~\ref{fig:phase3}d).
Finally, Fig.~\ref{fig:phase4} presents the phase-averaged statistics for case~4.
The phase-averaged streamwise velocity (Fig.~\ref{fig:phase4}a) is composed of a turbulent velocity base profile (Fig.~\ref{fig:umean}, dotted line) superimposed by the DTW-induced periodic streamwise velocity shown in Fig.~\ref{fig:phase4}(b).
In case~4, the phase-averaged streamwise velocity exhibits significantly lower values than in the relaminarization case (Fig.~\ref{fig:phase1}a).
As is the case in the latter instance, a backflow region is discernible around $\tilde{z}=R$.
Moreover, the periodic velocity components (Fig.~\ref{fig:phase4}b,c) as well as the periodic Reynolds shear stress (Fig.~\ref{fig:phase4}d) are comparable to the relaminarization case~1 (Fig.~\ref{fig:phase1}), which results in similar $\tilde{z}$-averaged periodic Reynolds shear stress profiles (cf. Fig.\ref{fig:rss}b).
For case~4, the turbulent Reynolds stresses (Fig.~\ref{fig:phase4}e,f,g,h) resemble the uncontrolled flow case, exhibiting minimal dependence on the streamwise phase coordinate $\tilde{z}$.
In the context of DTW-induced relaminarization, the flow transitions to a two-dimensional state, enabling the calculation of the two-dimensional stream function of the phase-averaged streamwise and radial velocity components
$\psi_{\langle u_z \rangle_{N_{\phi_z}\varphi t} \langle u_r \rangle_{N_{\phi_z}\varphi t}}$
that satisfies the condition
\begin{equation}
\begin{pmatrix} 
\partial \psi_{\langle u_z \rangle_{N_{\phi_z}\varphi t} \langle u_r \rangle_{N_{\phi_z}\varphi t}} / \partial z  \\ 
\partial \psi_{\langle u_z \rangle_{N_{\phi_z}\varphi t} \langle u_r \rangle_{N_{\phi_z}\varphi t}} / \partial r  
\end{pmatrix}
=
\begin{pmatrix} 
-\langle u_r \rangle_{N_{\phi_z}\varphi t}  \\ \langle u_z \rangle_{N_{\phi_z}\varphi t}
\end{pmatrix}
        \,.\label{eq:psi}
 \end{equation}
Fig.~\ref{fig:phasepsi}(a,b,c) presents iso-contours of $\psi_{\langle u_z \rangle_{N_{\phi_z}\varphi t} \langle u_r \rangle_{N_{\phi_z}\varphi t}}$, normalized with the corresponding laminar flow centerline velocity $U_{c,lam}$ and the pipe radius $R$.
These are compared between relaminarization cases for the different Reynolds numbers.
As shown in the left column of Fig.~\ref{fig:phasepsi}, the stream function demonstrates a considerable degree of independence from the Reynolds number when normalized by $U_{c,lam}$ and $R$. 
Nonetheless, close to the wall, wall-attached structures are visible, which appear not to scale with the pipe radius.
This is further investigated by taking into account the stream function of the periodic component of the streamwise and radial velocity, normalized by $U_{c,lam}$ and $\delta_\nu$ (Fig.~\ref{fig:phasepsi}, right column).
As demonstrated in Fig.~\ref{fig:phasepsi}(e,f,g), counter-rotating half vortices emerge, induced by the DTWs.
These vortices periodically accelerate and decelerate the flow, thereby generating the flow pattern depicted in Fig.~\ref{fig:phasepsi}(a,b,c).
Since for the selected cases, the wavelength was kept constant in wall units, i.e. $\lambda^+=360$, the elliptical vortices scale in wall units with a minor diameter of $d_z^+=1/2\lambda^+=180$ and a major diameter of $d_y^+\approx300$.
Consequently, the region of the direct influence of the traveling wave control appears to be approximately constant for the selected cases when normalized with the viscous length scale $\delta_\nu$.
The latter is consistent with the aforementioned scaling of the effective layer thickness $\eta$ (cf. Fig.~\ref{fig:etat}).\par
%
%
%
\section{Conclusion}
\label{sec:conclusion}
In the present study, we performed DNSs of turbulent pipe flow controlled by streamwise traveling waves of wall blowing and suction under the CPG condition at friction Reynolds numbers of $180 \le \mathrm{Re}_\tau \le 720$.
Through parametric studies involving variations in the traveling wave amplitude $a$, celerity $c$, and wavelength $\lambda$, we investigated the scaling of the drag reduction and the relaminarization for both UTWs ($c<0$) and DTWs ($c>0$).
As in channel flow studies~\cite{Min2006,Hoepffner2009,Mamori2014}, UTWs destabilize the flow, maintaining turbulence while reducing the friction drag below that of the corresponding laminar flow.
However, when considering the required actuation energy, most UTW cases, except for low-speed UTWs ($c=-0.1U_{c,lam}$), did not result in significant net energy savings compared to the uncontrolled case.
For standing waves ($c=0$), drag reduction and positive net energy savings are obtained for wave amplitudes up to $a=0.15U_{c,lam}$ and wavelengths of $\lambda^+=360$. 
Relaminarization and drag reduction were observed for all friction Reynolds numbers $180\le \mathrm{Re}_\tau\le 720$ for DTWs.

In agreement with earlier pipe flow studies at $\mathrm{Re}_\tau=110$~\cite{Koganezawa2019} and channel flow studies at $\mathrm{Re}_\tau=110$ and $\mathrm{Re}_\tau =300$~\cite{Mamori2014}, where no upper bounds were assessed for the wave amplitude $a$ and celerity $c$, a relatively large range of wave amplitudes $a\gtrsim 0.07U_{c,lam}$, celerities $c\gtrsim U_{c,lam}$, and wavelengths $90 \le \lambda^+\le 720$ lead to a remarkable reduction in friction drag depending on the Reynolds number.
Nonetheless, only a subset of these wave parameters prompt a positive net energy savings rate $S$.
Similar to the drag reduction rate, the wave amplitude and celerity leading to substantial net energy savings scale with the centerline velocity of the corresponding laminar flow, i.e. $0.067 \lesssim  a/U_{c,lam} \lesssim 0.1$ and $c\approx U_{c,lam}$, whereas the wavelength scales in wall units, $\lambda^+=360$.
The effective layer thickness $\eta$ that leads to significant drag reduction, which is linearly dependent on the traveling wave time period $T$, is in the range of $1.4 \lesssim \eta^+ \lesssim 11.5$ for $180 \le \mathrm{Re}_\tau \le 540$ and matches earlier results of $1 \lesssim \eta^+ \lesssim 10$ for $\mathrm{Re}_\tau = 110$~\cite{Koganezawa2019}.
However, the present study reveals that only effective layer thicknesses of $3.8 \lesssim \eta^+ \lesssim 5.7$ at a constant traveling wave period of $T=360\delta_\nu/U_{c,lam}$ generate  notable net energy savings.
While the maximum achievable drag reduction rate $R_D$ increases with the Reynolds number up to approximately $95\%$ at $\mathrm{Re}_\tau=540$, the net energy savings rate $S$ seems to settle at a value of $S\approx 90\%$ at $\mathrm{Re}_\tau\ge360$.
In the cases of relaminarization, the DTWs dampen the near-wall turbulence to zero within three dimensionless time units $D/u_\tau$, whereas the acceleration to the terminal velocity takes more than $65D/u_\tau$, relatively independent of the Reynolds number. 
The decay of the TKE can be modeled employing an exponential decay law, $k(t)\approx k(t=0)\exp{(-t/\tau)}$ with $\tau=0.55D/u_\tau$.
We explored the scaling of the mean velocity profile and the Reynolds stress components using time-, space-, and phase-averaged turbulence statistics. 
For the relaminarization cases, DTWs produce a backflow region close to the pipe wall with a negative velocity minimum at a constant radius of $r\approx 0.98R$.
This partial blockage of the pipe cross-section occasions a smaller mean velocity profile than the corresponding laminar Hagen-Poiseuille profile.
Nevertheless, more than $97\%$ of the theoretically achievable drag reduction rate is attained within the observed Reynolds number range of $180\le\mathrm{Re}_\tau\le540$.
Finally, the stream function of the phase-averaged periodic streamwise and radial velocity components for the relaminarization consists of half-vortical structures at the pipe wall.
The half-vortices induced by the DTWs and scaling in wall units are superimposed on the $z\varphi t$-averaged velocity profile, scaling with the pipe radius $R$. 
This results in mixed scaling of the stream function of the phase-averaged velocity.
To conclude, applying streamwise traveling waves of wall blowing and suction successfully relaminarizes initially turbulent pipe flow at friction Reynolds numbers up to $\mathrm{Re}_\tau=540$, all while saving net energy.
Furthermore, the observed scaling relations can be implemented to predict traveling wave flow control at even higher Reynolds numbers.\par
%

\backmatter

\bmhead{Supplementary information}

Supplementary movies showing the temporal evolution of volume-rendered flow quantities are electronically available.

\bmhead{Acknowledgements}

The authors sincerely appreciate the careful and conscientious proofreading provided by Mahsa Pakzad.

The authors gratefully acknowledge the scientific support and HPC resources provided by the German Aerospace Center (DLR). The HPC system CARO is partially funded by "Ministry of Science and Culture of Lower Saxony" and "Federal Ministry for Economic Affairs and Climate Action".

\section*{Declarations}


\begin{itemize}
\item \textbf{Competing interests:} The authors declare that they have no known financial or non-financial competing interests that could have appeared to influence the work reported in this manuscript.
\item \textbf{Data availability:} The data that support the findings of this study are available from the corresponding author upon reasonable request.
\item \textbf{Author contribution:} Both authors contributed to the study conception and design. CB performed the numerical simulations, data analysis, and drafted the initial manuscript. CW provided resources and critical revision of the manuscript. Both authors read and approved the final version of the manuscript.
\end{itemize}


%
%

%
\begin{appendices}

\section{Grid and domain length convergence}\label{secA1}
Here, we examine how sensitive the results are to grid resolution and domain length $L$.
For case~1 (DTWs and relaminarization), we performed two additional simulations: One with twice the grid resolution of the baseline case (case~1b) and another one with an increased domain length of $168R$ instead of $20R$ (case~1c).
For case~3 with UTWs, we performed one additional simulation with twice grid resolution of the baseline case (case~3b).
The additional cases, in conjunction with their respective baseline cases, are illustrated in  Table~\ref{tab:casesapp}.\par
\begin{figure}[h!] 
\centering
\includegraphics[width=\linewidth]{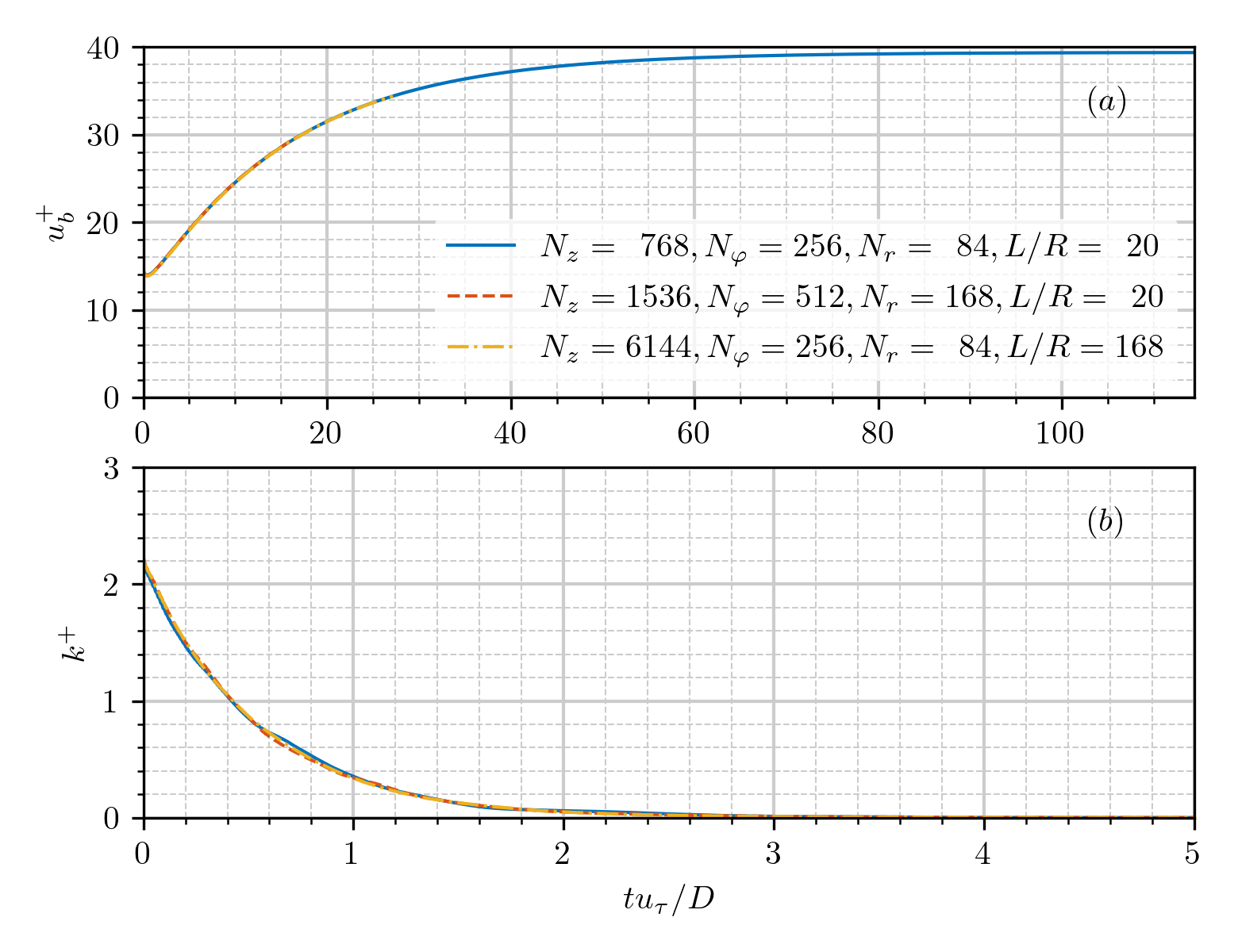}
\caption{Grid and domain length dependency of the bulk flow rate $u_b^+$ (a) and the turbulent kinetic energy $k^+$ (b) for $\mathrm{Re}_\tau=180$, $a^+=6$, $\lambda^+=360$, and $c^+=90$.}
\label{fig:gridstudy}
\end{figure}

\begin{table}[h] 
\caption{Grid and domain length convergence cases: $\mathrm{Re}_\tau=$ is the friction Reynolds number, $L/R$ is the domain length, $a^+$, $\lambda^+$, and $c^+$ are the amplitude, the wavelength, and the celerity of the traveling wave boundary condition in wall units. $N_z , N_{\phi}$, and $N_r$ are the number of grid points with respect to the axial, azimuthal, and radial direction, respectively. $\Delta{z^+}$, streamwise grid spacing; $R^+\Delta{\varphi}$ azimuthal grid spacing at the wall; $\Delta{r^+}$, minimal and maximal radial grid spacing.}
\begin{tabularx}{\linewidth}{lXccccccccc} 
\hline\hline
case  & $\mathrm{Re}_\tau$  & $L/R$    & $a^+$     & $\lambda^+$   & $c^+$ & $N_z\times N_\varphi \times N_r$ & $\Delta z^+$ & $R^+\Delta \varphi$ & $\Delta r^+$   \\
\hline
1    &180    &  20                  &  6        & 360           & 90     &768x256x84&4.7&4.4&0.31\dots 4.4  \\
1b   &180   &  20                  &  6        & 360           & 90     &1536x512x168&2.3&2.2& 0.15\dots 2.2 \\
1c   &180   &  168                  &  6        & 360           & 90     &6144x256x84&4.9&4.4&0.31\dots 4.4  \\
3    &180   &  20                  &  27       & 360           & -9     &768x256x84&4.7&4.4&0.31\dots 4.4   \\
3b   &180   &  20                  &  27       & 360           & -9     &1536x512x168&2.3&2.2& 0.15\dots 2.2   \\
\hline\hline
\label{tab:casesapp}
\end{tabularx}
\end{table}
Fig.~\ref{fig:gridstudy} displays time series of the bulk flow rate $u_b^+$ (a) and the turbulent kinetic energy $k^+$ (b).
Fig.~\ref{fig:gridstudy}(a) shows a collapse of the curves obtained from case~1, 1b, and~1c, thereby indicating that the bulk flow rate is robust with respect to an increase in grid resolution and an increase in domain length.
Note that the simulations with the increased grid resolution (case~1b) and with the increased domain length (case~1c) did not run as long as the baseline case~1.
The collapse of the curves also demonstrates the robustness of the flow control method with respect to different initial turbulent fields.
This is because both the refined grid and extended domain cases start from different initial fields than the baseline case.
This also causes the slight deviations in TKE time series of cases~1b and~1c compared to baseline case~1, as seen in Fig.~\ref{fig:gridstudy}(b).
Because the initial turbulent fields differ, the initial level and distribution of TKE differ, causing the TKE time series  not to collapse exactly.
Nevertheless, the relatively good collapse of the TKE time series proves that the the grid resolution and domain length of the baseline are sufficient.\par
\begin{figure}[h!] 
\centering
\includegraphics[width=\linewidth]{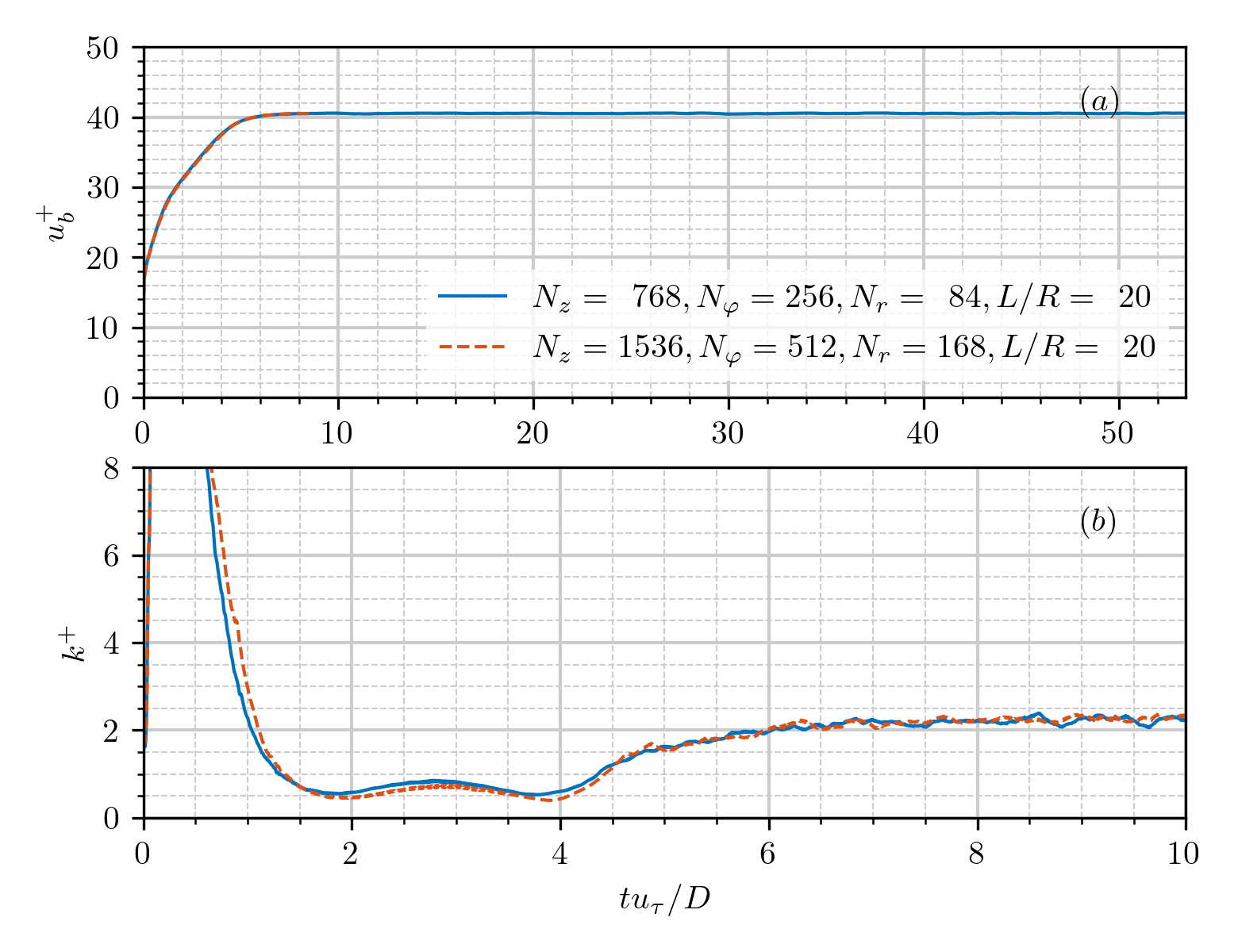}
\caption{Grid and domain length dependency of the bulk flow rate $u_b^+$ (a) and the turbulent kinetic energy $k^+$ (b) for $\mathrm{Re}_\tau=180$, $a^+=27$, $\lambda^+=360$, and $c^+=-9$.}
\label{fig:gridstudy2}
\end{figure}
\begin{figure} 
\centering
\includegraphics[]{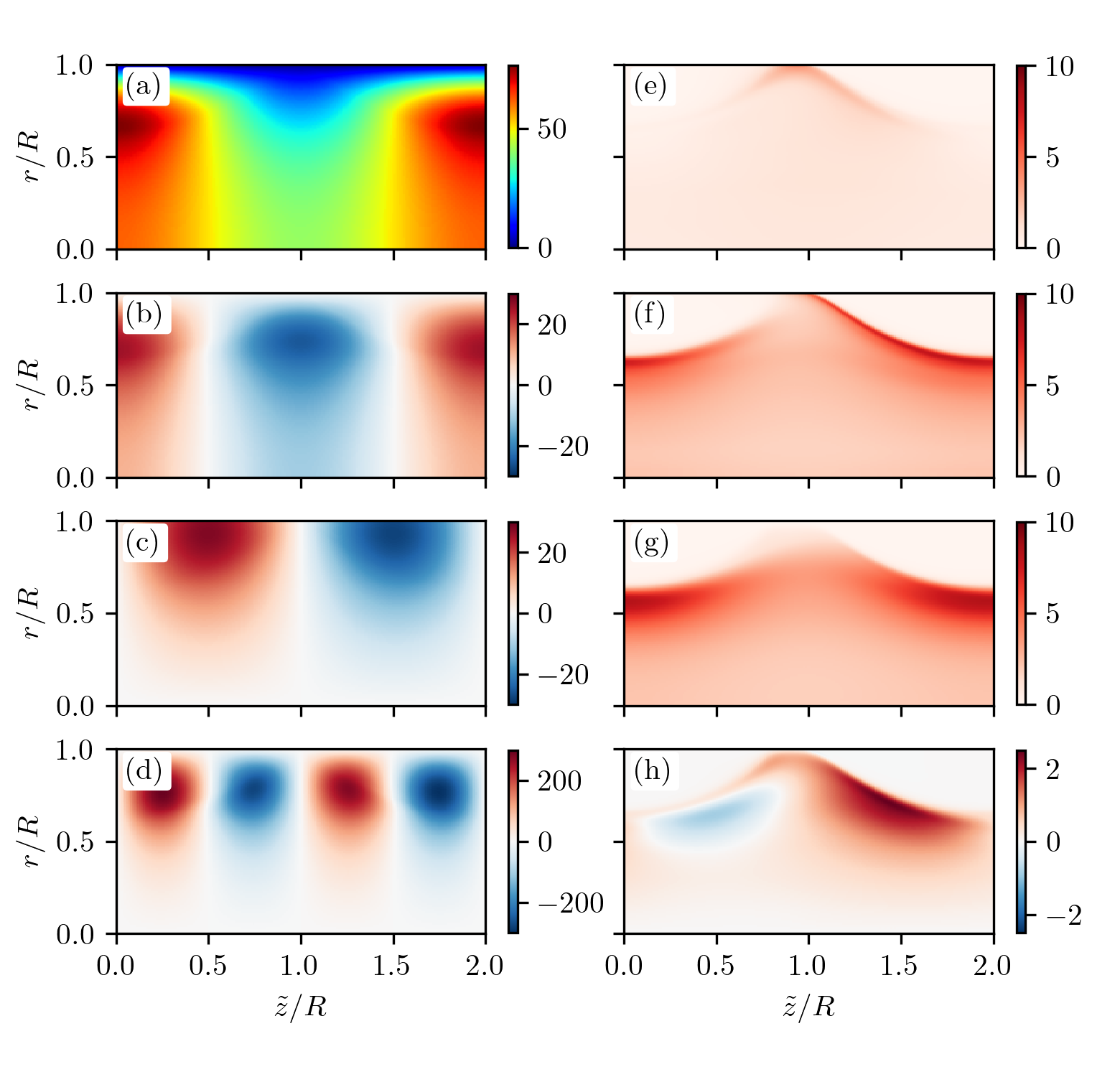}
\caption{Phase-averaged statistics for case 3 ($a^+=27$, $\lambda^+=360$, $c^+=-9$). 
(a) Phase-averaged streamwise velocity $\langle u_z\rangle_{N_{\phi_z}\varphi t}$; 
(b) periodic component of the streamwise velocity $\tilde{u}_z$; 
(c) phase-averaged radial velocity $\langle u_r\rangle_{N_{\phi_z}\varphi t}=\tilde{u}_r$; 
(d) periodic Reynolds shear stress $\tilde{u}_z \tilde{u}_r$;
(e) turbulent streamwise Reynolds stress $\langle u_z^{\prime\prime} u_z^{\prime\prime} \rangle_{N_{\phi_z \varphi t}}$
(f) turbulent azimuthal Reynolds stress $\langle u_\varphi^{\prime\prime} u_\varphi^{\prime\prime} \rangle_{N_{\phi_z \varphi t}}$
(g) turbulent radial Reynolds stress $\langle u_r^{\prime\prime} u_r^{\prime\prime} \rangle_{N_{\phi_z \varphi t}}$
(h) turbulent Reynolds shear stress $\langle u_z^{\prime\prime} u_r^{\prime\prime} \rangle_{N_{\phi_z \varphi t}}$.
}
\label{fig:phase3fine}
\end{figure}

For UTWs, Fig.~\ref{fig:gridstudy2} displays time series of the bulk flow rate $u_b^+$ (a) and the turbulent kinetic energy $k^+$ (b) of the cases~3 and~3b.
For the UTW case as well, the bulk flow rate time series of the baseline case and the finer grid cases collapse nicely (Fig.~\ref{fig:gridstudy2}a).
However, as for the the DTW case, the TKE time series do not collapse exactly and the deviation during the initial transient phase appears to larger than in the DTW case (Fig.~\ref{fig:gridstudy2}b).
This might be due to differences in the initial fields, which are amplified during the transitional phase when turbulent motions become most intense.
After the initial phase ($t\approx6D/u_\tau$), however,  the curves align again, fluctuating around the same mean value.
To further verify the validity of the baseline case grid resolution for UTWs, we examine the phase-averaged statistics of the refined case~3b in Fig.~\ref{fig:phase3fine}.
Comparing the latter with the phase-averaged statistics of the coarser baseline case~3 in Fig.~\ref{fig:phase3} reveals significant alignment for all depicted quantities.
Therefore, the grid resolutions and computational domain lengths for the different simulations given in Table~\ref{tab:cases}, which were been estimated and validated for fully-developed turbulent pipe flow, are also sufficient for the case of UTW- and DTW-controlled turbulent pipe flow.

\end{appendices}

\bibliography{references}%

\end{document}